\documentclass[letterpaper,11pt]{article}
\usepackage[margin=1in]{geometry}

\usepackage{amsmath}
\usepackage{graphicx}
\usepackage[pagebackref=true]{hyperref}
\usepackage[utf8]{inputenc}
\usepackage{natbib}
\usepackage{xcolor}
\usepackage{ifthen}
\usepackage[normalem]{ulem}
\usepackage{svg}
\usepackage{subfiles}
\usepackage{xspace}
\usepackage{cleveref}
\usepackage{wrapfig}
\usepackage{dialogue}
\usepackage[most]{tcolorbox}
\usepackage{subcaption}
\usepackage{longtable}

\usepackage{booktabs}
\usepackage{multirow}
\usepackage{titling}
\usepackage{arydshln}
\usepackage{pifont}

\usepackage{xcolor}

\definecolor{annotcolo}{rgb}{0.1,0.1,0.9}

\newcommand{\smalltextsc}[1]{\textsc{\small #1}}

\newcommand{\livecodebench}{\smalltextsc{LiveCodeBench}}
\newcommand{\lcb}{\smalltextsc{LCB}}

\newcommand{\llm}{\smalltextsc{LLM}\xspace}
\newcommand{\llms}{\smalltextsc{LLMs}\xspace}
\newcommand{\chainot}{\smalltextsc{COT}\xspace}
\newcommand{\python}{\smalltextsc{Python}\xspace}

\newcommand{\openai}{\smalltextsc{OpenAI}}
\newcommand{\gpts}{\smalltextsc{GPTs}\xspace}

\newcommand{\gptthreefiveturbo}{\smalltextsc{GPT-3.5-turbo}\xspace}
\newcommand{\gptfour}{\smalltextsc{GPT-4}\xspace}
\newcommand{\gptfourturbo}{\smalltextsc{GPT-4-Turbo}\xspace}
\newcommand{\gptfouro}{\smalltextsc{GPT-4-O}\xspace}

\newcommand{\claude}{\smalltextsc{Claude}\xspace}
\newcommand{\claudes}{\smalltextsc{Claudes}\xspace}
\newcommand{\claudetwo}{\smalltextsc{Claude-2}\xspace}
\newcommand{\claudethrees}{\smalltextsc{Claude-3s}\xspace}
\newcommand{\claudeinstantone}{\smalltextsc{Claude-Ins-1}\xspace}
\newcommand{\claudeopus}{\smalltextsc{Claude-3-Opus}\xspace}
\newcommand{\claudesonnet}{\smalltextsc{Claude-3-Sonnet}\xspace}

\newcommand{\gemini}{\smalltextsc{Gemini}\xspace}
\newcommand{\geminis}{\smalltextsc{Geminis}\xspace}
\newcommand{\geminipro}{\smalltextsc{Gemini-Pro}\xspace}
\newcommand{\geminiproonefive}{\smalltextsc{Gemini-Pro-1.5}\xspace}
\newcommand{\geminiflash}{\smalltextsc{Gemini-Flash}\xspace}

\newcommand{\cllamas}{\smalltextsc{CodeLLaMas}\xspace}
\newcommand{\cllama}{\smalltextsc{CodeLLaMa}\xspace}

\newcommand{\cllamabase}{\smalltextsc{CodeLLaMa-Base}\xspace}

\newcommand{\cllamabaseB}[1]{\smalltextsc{CL-Base-#1B}\xspace}
\newcommand{\cllamaBList}{\smalltextsc{CL-Ins-\{7, 13, 34\}B}\xspace}
\newcommand{\cllamabaseBList}{\smalltextsc{CL-Base-\{7, 13, 34\}B}\xspace}

\newcommand{\llamas}{\smalltextsc{LLaMa-3s}\xspace}

\newcommand{\llamabase}{\smalltextsc{L3-Base}}
\newcommand{\llamabaselist}{\smalltextsc{L3-Base-\{7, 70\}B}\xspace}
\newcommand{\llamainslist}{\smalltextsc{L3-Ins-\{7, 70\}B}\xspace}
\newcommand{\llamaB}[1]{\smalltextsc{L3-Ins-#1B}\xspace}

\newcommand{\wizardlong}{\smalltextsc{WizardCoder}\xspace}

\newcommand{\phind}{\smalltextsc{Phind-34B}\xspace}
\newcommand{\magicoders}{\smalltextsc{MagiCoders}\xspace}
\newcommand{\magicoder}[1]{\smalltextsc{MC-#1B}\xspace}
\newcommand{\magicoderBLit}{\smalltextsc{MC-\{6.7, 7\}B}\xspace}

\newcommand{\deepseek}{\smalltextsc{DeepSeek}\xspace}
\newcommand{\deepseekbase}{\smalltextsc{DeepSeek-Base}\xspace}
\newcommand{\deepseekinstruct}{\smalltextsc{DeepSeek-Instruct}\xspace}
\newcommand{\deepseeks}{\smalltextsc{DeepSeeks}\xspace}
\newcommand{\deepseekcode}{\smalltextsc{DS}\xspace}

\newcommand{\deepseekcodeB}[1]{\smalltextsc{DS-Ins-#1B}\xspace}
\newcommand{\deepseekbaseB}[1]{\smalltextsc{DS-Base-#1B}\xspace}
\newcommand{\deepseekcodeBList}{\smalltextsc{DS-Ins-\{1.3, 6.7, 33\}B}\xspace}
\newcommand{\deepseekcodebaseBList}{\smalltextsc{DS-Base-\{1.3, 6.7, 33\}B}\xspace}

\newcommand{\codeqwen}{\smalltextsc{CodeQwen}\xspace}

\newcommand{\mixtral}{\smalltextsc{Mixtral}\xspace}
\newcommand{\codestral}{\smalltextsc{Codestral}\xspace}
\newcommand{\mistrallarge}{\smalltextsc{Mistral-L}\xspace}
\newcommand{\mistral}{\smalltextsc{Mistral}\xspace}

\newcommand{\starcoder}{\smalltextsc{StarCoder}\xspace}
\newcommand{\starcodertwo}{\smalltextsc{StarCoder2}\xspace}
\newcommand{\starcodertwobase}{\smalltextsc{StarCoder2-Base}\xspace}
\newcommand{\sctwobaseB}[1]{\smalltextsc{SC2-Base-#1B}\xspace}
\newcommand{\sctwobaseBList}{\smalltextsc{SC2-Base-\{3,7,15\}B}\xspace}

\newcommand{\commandrplus}{\smalltextsc{CMD-R+}\xspace}

\newcommand{\apps}{\smalltextsc{APPS}\xspace}
\newcommand{\contests}{\smalltextsc{Code-Contests}\xspace}
\newcommand{\humaneval}{\smalltextsc{HumanEval}}
\newcommand{\humanevalplus}{\smalltextsc{HumanEval+}}
\newcommand{\mbpp}{\smalltextsc{MBPP}}
\newcommand{\cruxeval}{\smalltextsc{CRUXEval}}
\newcommand{\codet}{\smalltextsc{CodeT}}

\newcommand{\leetcode}{\smalltextsc{LeetCode}\xspace}
\newcommand{\atcoder}{\smalltextsc{AtCoder}\xspace}
\newcommand{\codeforces}{\smalltextsc{CodeForces}\xspace}
\newcommand{\html}{\smalltextsc{HTML}\xspace}

\newcommand{\easy}{\smalltextsc{Easy}\xspace}
\newcommand{\medium}{\smalltextsc{Medium}\xspace}
\newcommand{\hard}{\smalltextsc{Hard}\xspace}

\newcommand{\problem}{\smalltextsc{$P$}\xspace}
\newcommand{\solution}{\smalltextsc{$S$}\xspace}
\newcommand{\tests}{\smalltextsc{$T$}\xspace}

\newcommand{\contestdate}{\smalltextsc{$D$}\xspace}

\newcommand{\passmetric}[1]{\smalltextsc{Pass@}#1\xspace}

\definecolor{codegreen}{rgb}{0,0.6,0}
\lstdefinestyle{mystyle}{  
    commentstyle=\color{codegreen},
    keywordstyle=\color{blue},
    basicstyle=\ttfamily\small,
    breakatwhitespace=false,        
    breaklines=true,                 
    captionpos=b,                    
    keepspaces=true,                 
    showspaces=false,                
    showstringspaces=false,
    showtabs=false,                  
    tabsize=2
}
\newcommand{\smallsim}{\smallsym{\mathrel}{\sim}}

\makeatletter
\newcommand{\smallsym}[2]{#1{\mathpalette\make@small@sym{#2}}}
\newcommand{\make@small@sym}[2]{%
  \vcenter{\hbox{$\m@th\downgrade@style#1#2$}}%
}
\newcommand{\downgrade@style}[1]{%
  \ifx#1\displaystyle\scriptstyle\else
    \ifx#1\textstyle\scriptstyle\else
      \scriptscriptstyle
  \fi\fi
}
\makeatother

\definecolor{RoyalBlue}{rgb}{0.25, 0.41, 0.88}
\definecolor{brickred}{rgb}{0.8, 0.25, 0.33}
\definecolor{forestgreen}{rgb}{0.13, 0.55, 0.13}

\title{LiveCodeBench: Holistic and Contamination Free Evaluation of Large Language Models for Code}

\author{Naman Jain\textsuperscript{\textdagger} \quad \quad 
King Han\textsuperscript{\textdagger} \quad \quad 
Alex Gu\textsuperscript{*} \textsuperscript{\textdollar} \quad \quad 
Wen-Ding Li\textsuperscript{*}\textsuperscript{\textdaggerdbl} \quad  \\ \\
Fanjia Yan\textsuperscript{*}\textsuperscript{\textdagger} \quad \quad 
Tianjun Zhang\textsuperscript{*}\textsuperscript{\textdagger} 
\quad \quad 
Sida I. Wang
\quad \\ \\
Armando Solar-Lezama\textsuperscript{\textdollar} \quad \quad 
Koushik Sen\textsuperscript{\textdagger} \quad \quad 
Ion Stoica\textsuperscript{\textdagger} \\
{ }\\
\textsuperscript{\textdagger} UC Berkeley \quad \textsuperscript{\textdollar} MIT \quad \textsuperscript{\textdaggerdbl} Cornell 
\\
{ }\\
Website: \url{https://livecodebench.github.io/} \\
{ }\\
\texttt{\{naman\_jain,kingh0730,fanjiayan,tianjunz,ksen,istoica\}@berkeley.edu}\\
\texttt{gua@mit.edu} \quad \quad \texttt{asolar@csail.mit.edu} \quad \quad \texttt{wl678@cornell.edu} \quad 
} 

\renewcommand*\backref[1]{\ifx#1\relax \else (Cited on pg. #1) \fi}

\date{} 
\begin{document}

\maketitle

\noindent %
\begin{abstract}
\noindent Large Language Models (\llms{}) applied to code-related applications have emerged as a prominent field, attracting significant interest from both academia and industry. 
However, as new and improved \llms{} are developed, existing evaluation benchmarks (e.g., \humaneval{}, \mbpp{}) are no longer sufficient for assessing their capabilities.
In this work, we propose \livecodebench{}, a comprehensive and contamination-free evaluation of \llms{} for code, which collects \textit{new} problems over time from contests across three competition platforms, namely \leetcode{}, \atcoder{}, and \codeforces{}.
Notably, our benchmark also focuses on a broader range of code-related capabilities, such as self-repair, code execution, and test output prediction, beyond just code generation.
Currently, \livecodebench{} hosts over five hundred coding problems that were published between May 2023 and May 2024.
We have evaluated $18$ base \llms{} and $34$ instruction-tuned \llms{} on \livecodebench{}.
We present empirical findings on contamination, holistic performance comparisons, potential overfitting in existing benchmarks as well as individual model comparisons.
We will release all prompts and model completions for further community analysis, along with a general toolkit for adding new scenarios and models.
\end{abstract}

\clearpage
\renewcommand{\baselinestretch}{1.2}\normalsize 
\addtocontents{toc}{\protect\setcounter{tocdepth}{-1}}
\addtocontents{toc}{\protect\setcounter{tocdepth}{2}}
\renewcommand{\baselinestretch}{1.0}\normalsize 
\clearpage

\section{Introduction}
\label{sec:introduction}

Code has emerged as an important application area for \llms{}, with a proliferation of code-specific models 
\citep{chen2021evaluating,austin2021program,li2022competition,zhong2022codegen, allal2023santacoder, li2023starcoder, roziere2023code, deepseek-coder, luo2023wizardcoder, Phind, wei2023magicoder, ridnik2024code,starcoder2} and their applications across various domains and tasks such as program repair ~\citep{opencodeinterpreter, olausson2023demystifying}, optimization~\citep{madaan2023learning}, test generation~\citep{steenhoek2023reinforcement}, documentation generation~\citep{luo2024repoagent}, tool usage~\citep{patil2023gorilla, qin2024toolllm}, SQL~\citep{sun2023sql}, and more.
In contrast with these rapid advancements, evaluations have remained relatively stagnant, and current benchmarks like 
\humaneval{}, \mbpp{}, and \apps{} may paint a skewed or misleading picture.
Firstly, while coding is a multi-faceted skill, these benchmarks only focus on natural language-to-code tasks, thus overlooking broader code-related capabilities. 
Moreover, these benchmarks may be subject to potential contamination or overfitting, as benchmark samples are present in the training datasets.
\begin{figure}[!b]
    \centering
    \includegraphics[width=0.49\linewidth]{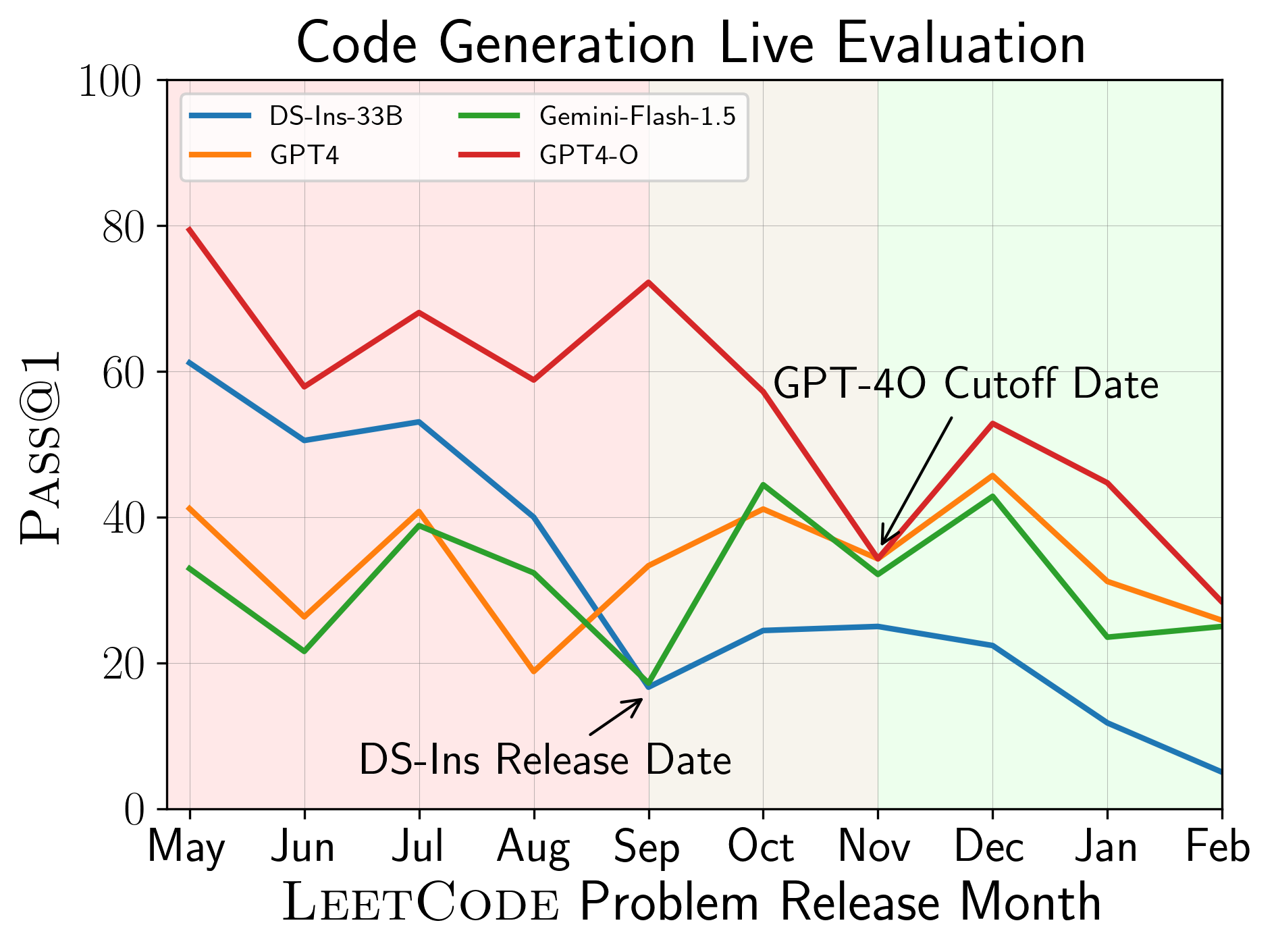}
    \hfil%
    \includegraphics[width=0.49\linewidth]{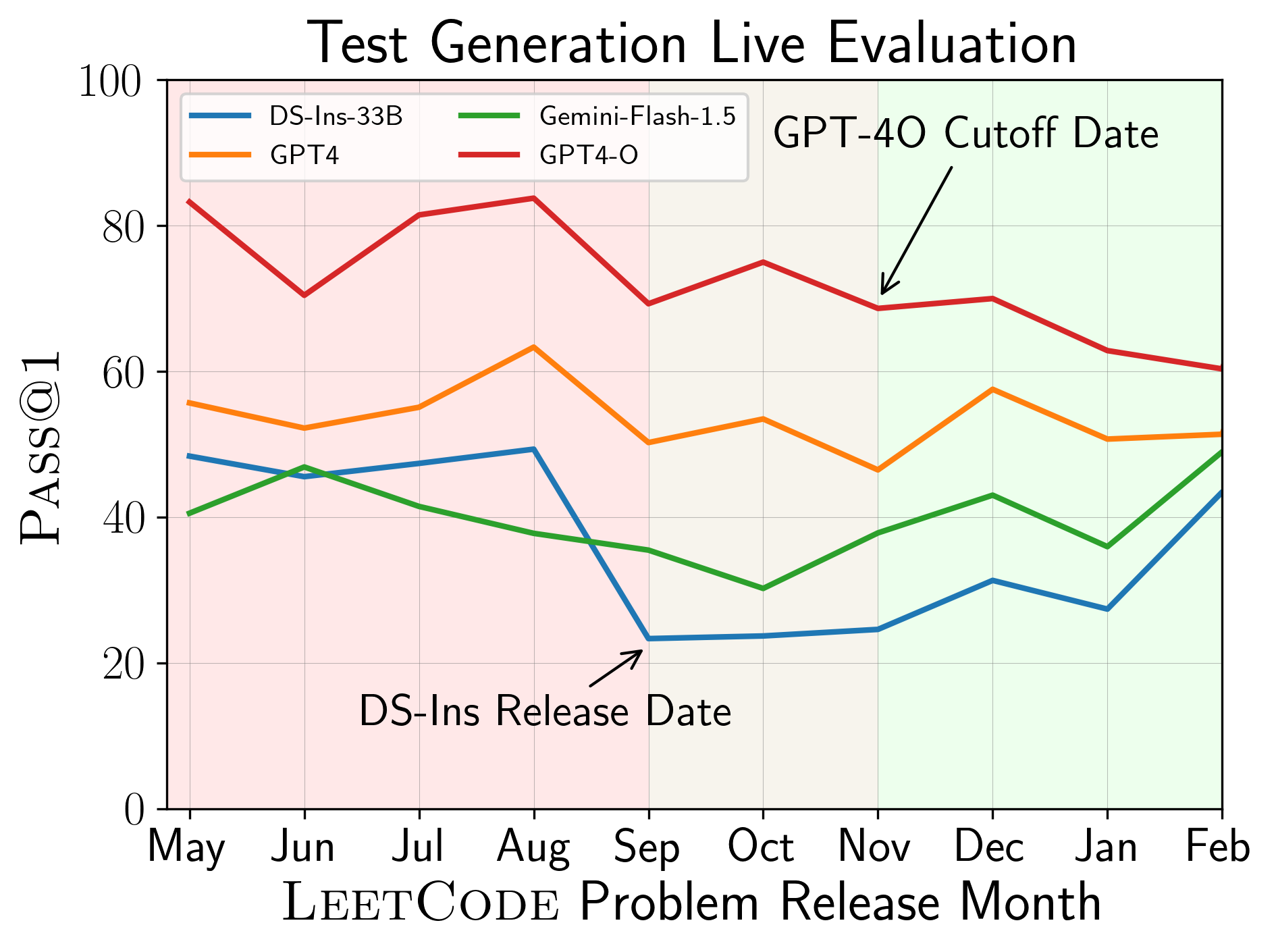}%
    \caption{
        \livecodebench{} comprises problems marked with release dates, allowing evaluations over different time windows.
        For newer models, we can detect and avoid contamination by only evaluating on time-windows after the model's cutoff date.
        The figures demonstrate the performance of models on code generation and test output prediction \livecodebench{} scenarios with \leetcode{} problems released across the months between May $2023$ and February $2024$. 
        Notice that \deepseekinstruct{} and \gptfouro{} perform considerably worse on problems released since September and November $2023$ (their release and cutoff dates respectively!) -- indicating potential contamination for the earlier problems. 
        Thus, while performing evaluations, we use the post-September/post-November time window (green) for fairly comparing these models. 
        }
    \label{fig:contamination_main}
\end{figure}

\begin{figure}[!b]
    \centering
    \includegraphics[width=0.47\linewidth]{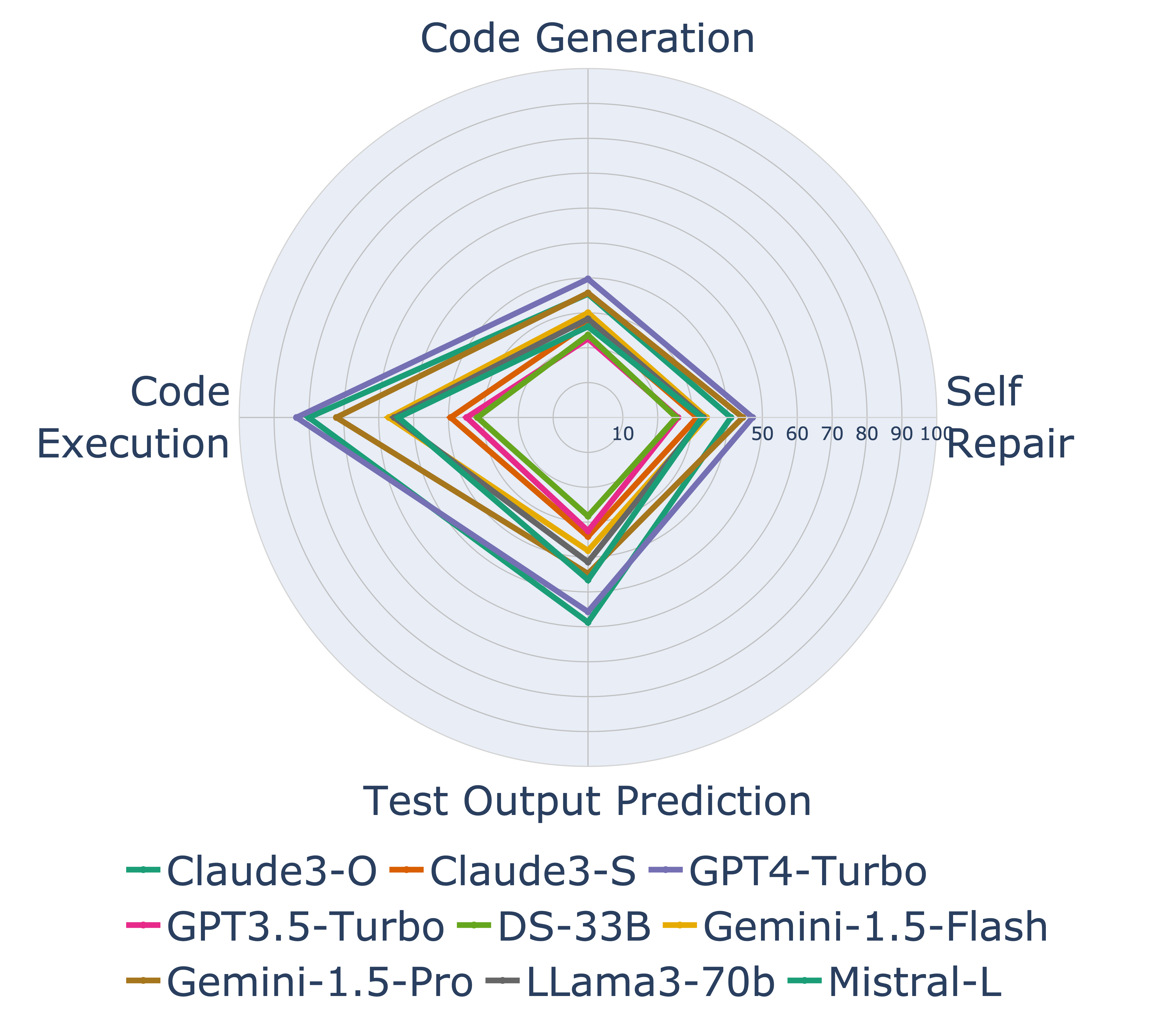}
    \hfil%
    \includegraphics[width=0.52\linewidth]{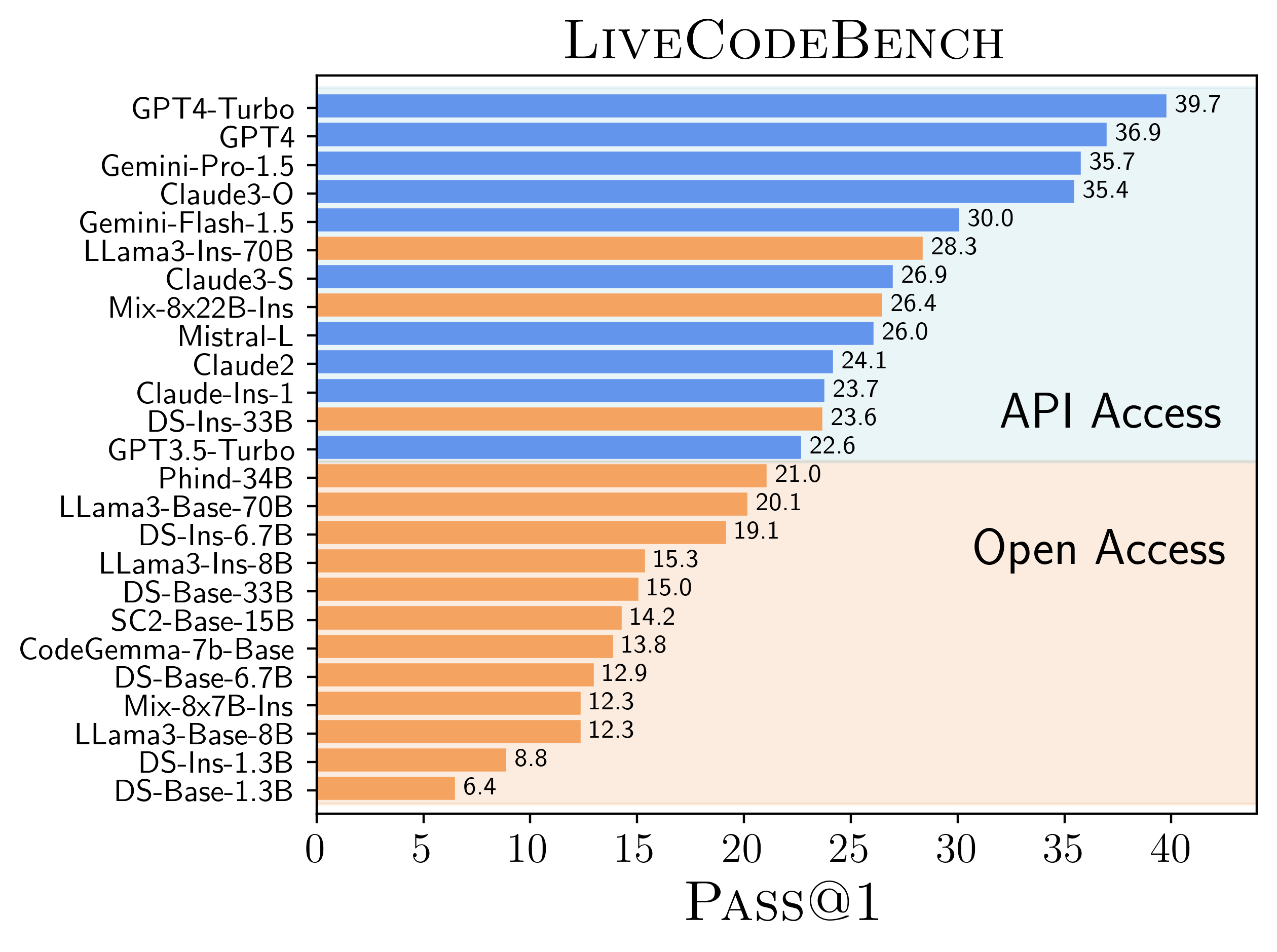}%
    \caption{
        \textbf{Left.} 
        We propose evaluating \llms{} across 
        scenarios capturing various coding-related capabilities.
        Specifically, we host four different scenarios, namely code generation, 
        self-repair, code execution, and test output prediction.
        The figure depicts various model performances across the four scenarios available in \livecodebench{} in a radial plot -- 
        highlighting how relative differences across models change across the scenarios.
        \textbf{Right.}
        Comparison of open access and (closed) API access models on \livecodebench{}-Easy 
        code generation scenario. We find that closed-access models consistently 
        outperform the open models with only strong instruction-tuned variants of $>30$\textsc{B}
        models (specifically \llamaB{70}, \mixtral and  \deepseekcodeB{33} models) crossing the performance gap.
        }
    \label{fig:all_tasks_radial}
\end{figure}

Motivated by these shortcomings, we introduce \livecodebench{}, 
a holistic and contamination-free benchmark for evaluating code capabilities. 
\livecodebench{} is built on the following principles:
\begin{enumerate}
\itemsep-0.1em 
    \item \textbf{Live updates to prevent contamination.} 
    \llms{} are trained on massive inscrutable corpora, and current benchmarks 
    suffer from the risk of data contamination as they could be included in those training datasets.
    While previous works have attempted decontamination using both exact and fuzzy matches~\citep{li2023starcoder, li2023textbooks}, it can be a non-trivial task ~\citep{gemini15} and can be evaded using simple strategies like rephrasing~\citep{yang2023rethinking}.
    Here, to prevent the risk of problem contamination, we use live updates, 
    that is evaluate models on \textit{new} problems. 
    Particularly, we collect problems from weekly contests on competition platforms 
    and tag them with a \textit{release date}. 
    Next, for newer models, we only consider problems released after the model's cutoff date to ensure that the model has not encountered the exact problem in the training dataset.
    In Figure~\ref{fig:contamination_main}, we find that the performance of the \deepseek{} model starkly 
    drops when evaluated on the \leetcode{} problems released after August $2023$.  
    Similarly, \gptfouro{} observes a drop in performance on \leetcode{} problems released since November 2023, its specified cutoff date.
    This indicates that these models are likely trained on the older \leetcode{} problems and time-segmented evaluations allow fair comparisions.
    
    \item \textbf{Holistic Evaluation.} 
    Current code evaluations primarily focus on natural language to code generation. 
    However, programming is a multi-faceted task that requires a variety of capabilities beyond those measured by code generation. 
    In \livecodebench{}, we evaluate code \llms{} on three additional scenarios, listed below. 
    \begin{itemize}
    \item \textbf{Self-Repair.} Fix an incorrect program from execution information, evaluating the ability to debug code from feedback. The model is given the natural language problem description, the incorrect program, the test case it fails on, and the execution feedback from that failure. The output should be a correct repaired program.
    \item \textbf{Code Execution.} ``Execute'' a program on an input, evaluating code comprehension ability. The model is given a program and an input, and the output should be the result.
    \item \textbf{Test Output Prediction.} Solve the natural language task on a specified input, evaluating the ability to generate testing outputs. The model is given the natural language problem description and an input, and the output should be the output for the problem.
    \end{itemize}
    Figure~\ref{fig:all_tasks_radial} (left) depicts performance on the different scenarios considered in \livecodebench{}. 
    \item \textbf{High-quality problems and tests.}
    High-quality problems and tests are crucial for reliable evaluation of \llms{}.
    However, prior works have revealed deficiencies in existing benchmarks. 
    \citep{liu2023evaluating} identified insufficient tests and ambiguous problem descriptions in \humaneval{}. They released \humanevalplus{}, a variant of the benchmark with more tests and sometimes saw up to an 8\% drop in performance. 
    Similarly, \citep{austin2021program} had to create a sanitized \mbpp{} subset to disambiguate problem descriptions.
    In \livecodebench{}, we source the problems from reputable competition websites whose quality is already validated
    by the platform users.
    In addition, for every problem, we provide a good number of tests (about $17$ on average) for meaningful and robust evaluations while still finishing quickly.
    
    \item \textbf{Balanced problem difficulty.} 
    Competition programming is challenging for even the best-performing \llms{}, and most of the current SoTA models achieve close to zero performance on a majority of problems.
    As a result, they can be unsuitable for meaningful comparing today's \llms{} because the variance in performances is low.
    Furthermore, the averaging of evaluation scores across problems with different difficulty levels artificially minimizes the differences between models.
    Therefore, we use problem difficulty ratings (sourced from the competition websites) for filtering the harder problems and 
    classifying problem difficulties to
    ensure balanced problem difficulty distribution and allow granular model comparisons.
\end{enumerate}

\noindent With these principles in mind, we build \livecodebench{}, a continuously updated benchmark that avoids data contamination.
Particularly, we have collected $511$ problems from contests across three competition platforms -- \leetcode{}, \atcoder{}, and \codeforces{} occurring from May $2023$ to the present (May $2024$) and use them to construct the different \livecodebench{} scenarios. 
\newline
\newline
\noindent \textbf{Empirical Findings.} 
We have evaluated $18$ base models and $34$ instruction-tuned models across different \livecodebench{} scenarios.
Below, we present the empirical findings from our evaluations, which have not been revealed in prior benchmarks.

\begin{enumerate}
\itemsep-0.1em 
    \item \textbf{Contamination.} 
    We observe a stark drop in the performance of \deepseek{}, \gptfouro{}, and \codestral{} on \leetcode{} problems released after Aug $2023$, Oct $2023$, and Jan $2024$ (Figure~\ref{fig:contamination_main}). These results highlight likely contamination in older problems and time-segmented evaluations 
    prove effective for performing fair comparisons.
    
    \item \textbf{Holistic Evaluation.} 
    Our evaluations reveal that model performances are correlated across tasks, but the relative differences do vary. 
    For example, in Figure~\ref{fig:all_tasks_radial}, the gap between open and closed models further increases on tasks like self-repair or test output prediction. 
    Similarly,  \claudeopus{} and \mistrallarge{} perform considerably better on code execution and test output prediction compared to code generation with \claudeopus{} surpassing \gptfour{} on the test output prediction. This highlights the importance of a holistic evaluation.
    
    \item \textbf{\humaneval{} Overfitting.} 
    Upon comparing \livecodebench{} with \humaneval{}, we find that models cluster into two groups, ones that perform well on both benchmarks and others that perform well on \humaneval{} but not on \livecodebench{} (see Figure~\ref{fig:he_vs_lcb}).
    The latter group primarily comprises fine-tuned open-access models while the former group comprises base models and closed models.
    This indicates that these models might be overfitting to \humaneval{}. 

    \item \textbf{Model Comparisons} (Figure \ref{fig:results_plots})
    \vspace{-6pt}
    \begin{enumerate}
    \itemsep-0.1em 
    \item Among the open access base models, we find that \llamabase{} and \deepseekbase{} models are the strongest, followed by \starcodertwobase{} and \cllamabase{}.
    \item Closed API models such as \gpts{}, \claude{}, and \gemini{} generally outperform open models (Figure~\ref{fig:all_tasks_radial}). The open models that close the gap are \llamaB{70}, \mixtral{}, and \deepseekcodeB{33} are instruction-tuned variants of large base models ($> 30$\textsc{B} parameters). 
    \item Existing benchmarks are insufficient at highlighting the gap between \gptfour{} and other models. 
    Particularly, smaller models achieve similar or often even better performance compared to \gptfour{}. 
    In \livecodebench{}, \gptfour{} (and \gptfourturbo{}) outperforms all other models (except \claudeopus{}) by a large margin in all scenarios.
    
    \end{enumerate}
    
\end{enumerate}

\noindent \textbf{Concurrent Work.} 
~\citep{huang2023competition} also evaluate \llms{} in a time-segmented manner. 
However, they only focus on \codeforces{} problems while we combine problems across platforms 
and additionally propose a holistic evaluation across multiple code-related scenarios. 
~\citep{li2023taco} propose a large dataset of competitive programming problems with additional generated tests but do not study contamination or tasks beyond generation.
~\cite{liu2024codemind} evaluate the code comprehension capabilities of \llms{} using execution. \citep{singhal2024nofuneval} also propose evaluating \llms{} on tasks beyond code generation, but they consider tasks that take into account the \textit{non-functional-correctness aspects} of programming.
\citep{deepseek-coder} also evaluate \deepseek{} on \leetcode{} problems and mention the possibility of problem contamination.



\section{Holistic Evaluation}
\label{sec:tasks}

Code capabilities of \llms{} are evaluated and compared using natural language to code generation tasks.
However, this only captures one dimension of code-related capabilities.
Indeed, real-world software engineering requires expertise in tasks beyond just \textit{generation}, such as synthesizing informative test cases, debugging incorrect code, understanding existing code, and writing documentation.
These tasks are not just additional bookkeeping; they are crucial parts of the software development process and contribute to improving the quality, maintainability, and reliability of the code~\citep{10.1145/1134285.1134288}.
This also applies to \llms{} and adopting similar workflows can enable the models to perform better code generation.
For example, AlphaCodium~\citep{ridnik2024code} is an intricate \llm{} pipeline for solving competition coding problems. 
By combining natural language reasoning, test case generation, code generation, and self-repair, they achieve significant improvements over a naive direct code generation baseline, showcasing the importance of these broader capabilities.
Motivated by this, we propose a more holistic evaluation of \llms{} in this work using a suite of evaluation setups that capture a broader range of code-related capabilities.

\noindent Specifically, we evaluate code \llms{} in four scenarios, namely code generation, self-repair, code execution, and test output prediction.
Our selection criterion was to pick settings that are useful components in code \llm{} workflows 
and in addition, have clear and automated evaluation metrics.

\vspace{1.5pt}
\noindent Following we describe each of these scenarios in detail.

\begin{figure*}[!tb]
    \centering
    \includegraphics[width=\textwidth]{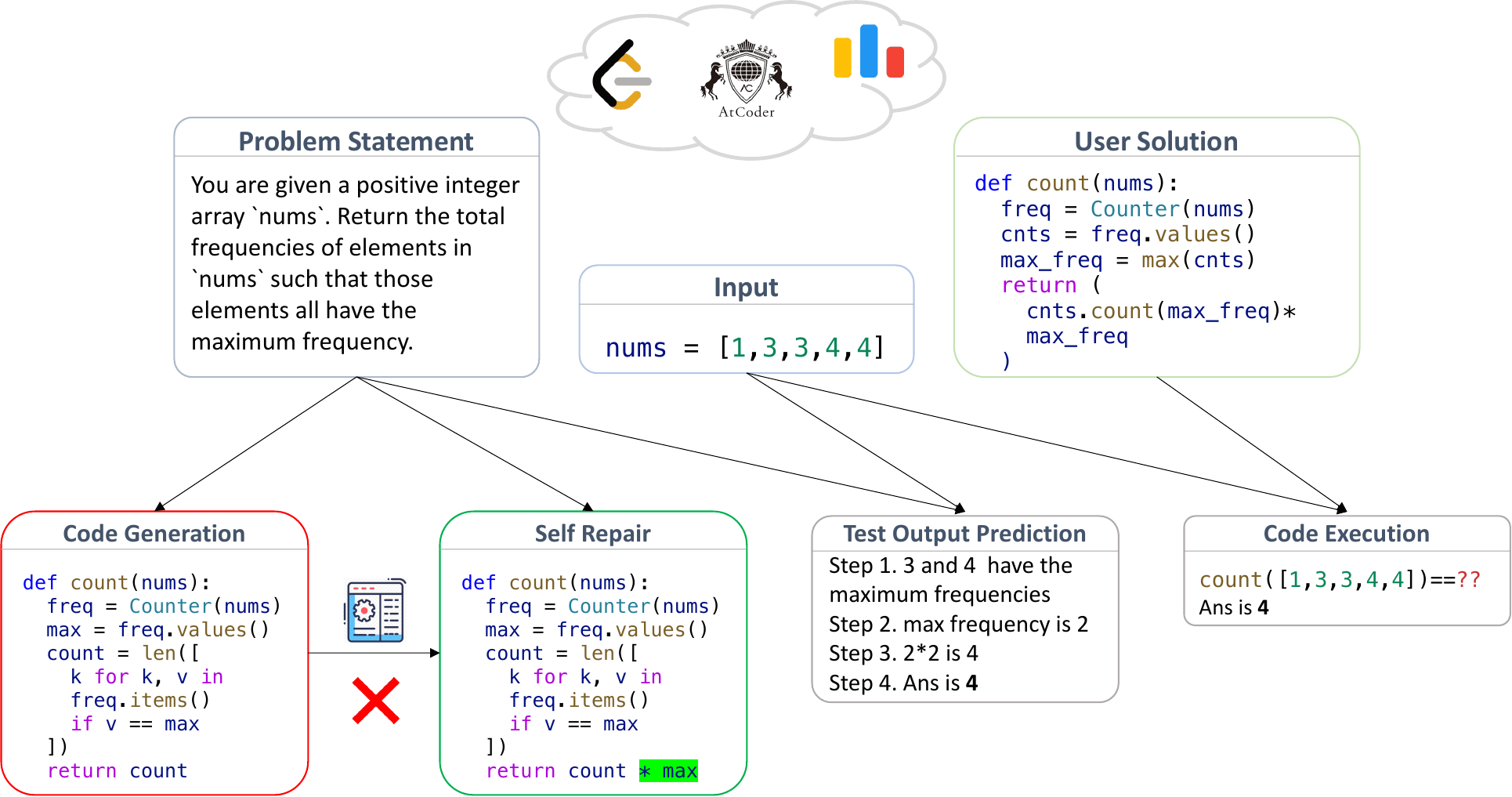}
    \caption{
        Overview of the different scenarios present in \livecodebench{}.
        Coding is multi-faceted and we propose evaluating \llms{} on a suite of evaluation setups that capture various coding-related capabilities.
        Specifically, beyond the standard code generation setting, we consider three additional scenarios, namely self-repair, code execution, and a newly introduced test output prediction task.
        }
    \label{fig:holistic_tasks}
    \vspace{-10pt}
\end{figure*}

\vspace{3pt}
\noindent \textbf{Code Generation.}
The code generation scenario follows the standard setup for generating code from natural language.
The model is given a problem statement, which includes a natural language description and example tests (input-output pairs), and is tasked with generating a correct solution.
The evaluation is performed based on functional correctness, using a set of \textit{unseen} test cases. 
We use the \passmetric{1} metric measured as the fraction of the problems for which the model was able to generate a program passing all tests.
Figure~\ref{fig:holistic_tasks} (left) provides an example of this scenario.

\vspace{3pt}
\noindent \textbf{Self Repair.}
The self-repair scenario is based on previous works that tested the self-repair capabilities of \llms{}~\citep{olausson2023demystifying,shinn2023reflexion,chen2023teaching}. 
Here, the model is given a problem statement from which it generates a candidate program (similar to the single-step code generation scenario above). 
However, in case of a mistake, the model is additionally provided with error feedback (either the exception message or a failing test case in case of incorrect code generation) and is tasked with generating the fixed solution.
Similar to the code generation scenario, the evaluation is performed via functional correctness on the final program, i.e. either the single-step correct generation or the attempted repair.
We use the  \passmetric{1} metric to measure the combined performance after the repair step.
Figure~\ref{fig:holistic_tasks} (mid-left) provides an example of this scenario.

\vspace{3pt}
\noindent \textbf{Code Execution.} 
The code execution scenario is based on the output prediction setup used in \cruxeval{} \citep{gu2024cruxeval}. 
The model is provided a program snippet consisting of a function (\texttt{f}) along with a test input to the program and is tasked with predicting the output of the program on the input test case.
The evaluation is performed via an execution-based correctness metric where the model generation is considered correct if \texttt{assert f(input) == generated\_output} passes. 
Figure~\ref{fig:holistic_tasks} (right) provides an example of the code execution scenario.

\vspace{3pt}
\noindent \textbf{Test Case Output Prediction.} 
Finally, we introduce a new task that is designed to study natural language reasoning and test generation. 
In this task, the model is given the problem statement along with a test case input, and it is tasked with generating the expected output for that input. 
This task follows a setup similar to the one used in \codet{}~\citep{chen2022codet}, where tests are generated solely from problem statements, without the need for the function's implementation.
A key difference is that we provide a fixed set of test inputs for each problem in our dataset, and the models are then prompted to only predict the expected output for those specific inputs. 
This approach allows for a straightforward evaluation of the test generation capabilities by avoiding test input prediction, a hard-to-evaluate task.
Figure~\ref{fig:holistic_tasks} (mid-right) provides an example of this scenario.
Finally, we would like to point out that \livecodebench{} also offers an extensible framework to add new scenarios in the future.
So other relevant settings like input generation, program summarization, optimization, etc. can be integrated with our setup.

\section{Benchmark Curation}
\label{sec:benchmark}
We curate our problems from three coding competition websites: \leetcode{}, \atcoder{}, and \codeforces{}. 
These websites periodically host contests containing problems that assess the coding and problem-solving skills of participants.
The problems consist of a natural language problem statement along with example input-output examples, and the goal is to write a program that passes a set of hidden tests.
Further, thousands of participants participate, solving these problems thus ensuring that the problems are vetted for clarity and correctness.

\subsection{Data Collection} 
We have written \html{} scrapers for each of the above websites to collect problems and the corresponding metadata.
To ensure quality and consistency, we parse mathematical formulas and exclude problems with images.
We also exclude problems that are not suitable for grading by input-output examples, such as those that accept multiple correct answers or require the construction of data structures.
Besides parsing the problem descriptions, we also collect associated ground truth solutions and test cases whenever directly available.
Thus for each problem, we collect tuples of  natural language problem statement \problem{}, 
test cases \tests{}, and ground truth solution \solution{}.
Finally, we associate the contest date \contestdate{} to mark the release date of each problem and 
use the collected attributes to construct problems for our four scenarios (detailed in Section~\ref{subsec:scenario-specific} ahead). 

\vspace{1pt}
\noindent \textbf{Scrolling through time.}
As noted, we associate the contest date \contestdate{} for each problem.
The release date allows us the measure the performance of \llms{} over different time windows by filtering problems 
based on whether the problem release date falls within a time window (referred to as ``scrolling'' through time).
This is crucial for evaluating and comparing models trained at different times. 
Specifically, for a new model and the corresponding cutoff date (normalized to the release date if the 
training cutoff date is not published), 
we can measure the performance of the model on benchmark problems released after the cutoff date.
We have developed a UI that allows comparing models on problems released during different time windows (shown in Figure~\ref{app:fig:scrolling}).

\vspace{1pt}
\noindent \textbf{Test collection.}
Tests are crucial for assessing the correctness of the generated outputs and are used in all four scenarios.
We collect tests available on platform websites whenever possible and use them for the benchmark.
Otherwise, following \citet{liu2023your}, we use a \llm{} (here \gptfourturbo{}) to generate tests for the problems.
A key difference between our test generation approach is that instead of generating inputs directly using the \llm{}, we construct generators that sample inputs based on the problem specifications using in context learning.
Details and examples of such input generators can be found in Section~\ref{appendix:input-generators}.
Finally, we collect a small fraction of failing tests from the platform for the more recent problems allowing more directed adversarial test collection.

\vspace{1pt}
\noindent \textbf{Problem difficulty.}
Competition programming has remained a challenge for \llms{}, with \gptfour{} achieving an average \codeforces{} rating (ELO) of 392, placing it in the bottom 5 percentile ~\citep{openai2023gpt}. 
This makes it difficult to compare \llms{}, as the variation in performance across models is low.
In \livecodebench{}, we collect problems of diverse difficulties as labeled in competition platforms, excluding problems that are rated above a certain threshold that are likely too difficult for even the best models\footnote{From our early explorations, we find {\scriptsize	\textsc{CodeForces}} problems being considerably more difficult than {\scriptsize \textsc{AtCoder}} and {\scriptsize \textsc{LeetCode}}  problems and thus focus primarily on the latter platforms.}.
Further, we use these ratings to classify problems as \easy{}, \medium{}, and \hard{} for more granular model comparisons.

\begin{table*}[!t]
    \centering
    \begin{tabular}{lcccccc}
    \hline
    Platform & Total Count & \#Easy & \#Medium & \#Hard & Average Tests \\
    \hline
    \lcb{} (May-end) & 511 & 182 & 206 & 123 & 17.0 \\
    \lcb{} (Sep-end) & 349 & 125 & 136 & 88 & 18.0 \\
    \hdashline
    \atcoder{} & 267 & 99 & 91 & 77 & 15.6 \\
    \leetcode{} & 235 & 79 & 113 & 43 & 19.0 \\
    \codeforces{} & 9 & 4 & 2 & 3 & 11.1 \\
    \hdashline
    \lcb{}-Easy & 182 & 182 & 0 & 0 & 16.1 \\
    \lcb{}-Medium & 206 & 0 & 206 & 0 & 17.4 \\
    \lcb{}-Hard & 123 & 0 & 0 & 123 & 18.0 \\
    \hline
\end{tabular}
    \caption{
        The statistics of problems collected in \livecodebench{} (\lcb{}). 
        We present the number of problems, their difficulty distributions and the average number of tests per problem.
        We present the results on the following subsets of \livecodebench{} (used throughout this manuscript) - (a) problems in the  May'23-May'24 and Sep'23-May'24 time windows, (b) problems sourced from the three platforms, and (c) problems in the \lcb{}-Easy, \lcb{}-Medium, and \lcb{}-Hard subsets.}
        \label{tab:platform_comparison}
\end{table*}

\subsection{Platform Specific Curation}
\label{subsec:platform-specific}
We describe the curation process for each platform.

\vspace{2pt}
\noindent \textbf{\leetcode{}.}
We collect problems from all weekly and biweekly contests on \leetcode{} that have taken place after April'23.
For each problem, we collect the problems, public tests, and user solutions.
The platform also provides a difficulty label for each problem which we use to tag the problems as \easy{}, \medium{}, and \hard{}.
Since \leetcode{} provides a starter code for each problem, we also collect it and provide it to the \llm{} in the STDIN format.
Since the hidden tests are not directly available, we use our generator-based test input generation approach (Section~\ref{appendix:input-generators}) and also collect the auto grader failing tests for some of the recent problems.

\vspace{2pt}
\noindent \textbf{\atcoder{}.}
We collect problems from the \texttt{abc} (beginner round) contests on \atcoder{} that have taken place after April'23. 
We deliberately avoid the more challenging \texttt{arc} and \texttt{agc} contests which are designed for more advanced Olympiad participants.
The problems are assigned numeric difficulty ratings, and we exclude \texttt{abc} problems with a rating of more than $500$.
We also use these numeric ratings to tag the problems as \easy{}, \medium{}, and \hard{}.
Specifically, we use the rating brackets $[0-200)$, $[200-400)$, and $[400-500]$ to perform the classification.
\atcoder{} provides public and hidden tests for each problem which we directly use in the benchmark.

\vspace{2pt}
\noindent \textbf{\codeforces{}.}
We have collected problems from the Division 3 and Division 4 contests on \codeforces{}.
Notably, we find that even with this filter, the problems are harder than the other two platforms.
\codeforces{} also provides difficulty ratings for the problems which we use to tag the problems as \easy{}, \medium{}, and \hard{} using the rating brackets $\{800\}$, $(800-1000]$, and $(1000-1300]$ respectively.
Due to the higher difficulty, we only consider a small fraction of problems from \codeforces{} and semi-automatically construct test case generators, as they do not provide complete tests on the platform (long tests are truncated).

\noindent Table~\ref*{tab:platform_comparison} provides various statistics about the problems that we have collected for \livecodebench{}. 

\subsection{Scenario-specific benchmark construction}
\label{subsec:scenario-specific}

\textbf{Code Generation and Self-Repair.}
We use the natural language problem statement as the problem statement for these scenarios. 
For \leetcode{}, as noted above, an additional starter code is provided for the functional input format.
For \atcoder{} and \codeforces{} problems, we use the standard input format (similar to ~\citet{hendrycks2021measuring}). 
The collected or generated tests are then used to evaluate the correctness of the generated programs.
Our final dataset consists of $511$ problem instances across the three platforms. 

\vspace{2pt}
\noindent \textbf{Code Execution.} 
We draw inspiration from the benchmark creation procedure used in ~\citet{gu2024cruxeval}. 
First, we collect a large pool of $\smallsim 2000$ \textit{correct, human-submitted solutions} from the \leetcode{} subset. 
However, many of these programs have multiple nested loops, complex numerical computations, and a large number of execution steps. 
Therefore, we apply compile-time and run-time filters to ensure samples are reasonable, and we double-check this with a manual inspection. 
More details on the filtering criteria and statistics of the dataset can be found in Appendix~\ref{appendix:dataset-execution}. 
Our final dataset consists of $479$ samples from $85$ problems.

\vspace{2pt}
\noindent \textbf{Test Case Output Prediction.} 
We use the natural language problem statement from the \leetcode{} platform and the example test inputs to construct our test case output prediction dataset.
Since the example test inputs in the problems are reasonable test cases for humans to reason about and understand the problems, they also serve as ideal test inputs for \llms{} to process.
Our final dataset consists of $442$ problem instances from a total of $181$ \leetcode{} problems.

\section{Experiment Setup}
\label{sec:setup}
We describe the experimental setup in this section.
First, we provide the common setup across the scenarios, followed by the scenario-specific setups in Section~\ref{subsec:scenario_specific_setup}.

\vspace{2pt}
\noindent \textbf{Models.} 
We evaluate $52$ models across various sizes, ranging from $1.3$B to $70$B, including base models, instruction models, and both open and closed models. 
Our experiments include models from different classes, such as \gpts{} (\gptthreefiveturbo{}, \gptfour{}, \gptfourturbo{},\gptfouro{}), \claudes{} (\claudeinstantone{}, \claudetwo{}, \claudethrees{}), \geminis{}(\geminipro{}, \geminiflash{}), \mistral{} among closed-access and \llamas{}(\llamabaselist{}, \llamainslist{}), \deepseeks{} (\deepseekcodebaseBList{}, \deepseekcodeBList{}), \cllamas{} (\cllamaBList{}, \cllamabaseBList{}), \starcodertwo{} (\sctwobaseBList{}), \codeqwen{} among open.
Additionally, we also include fine-tuned models \phind{} from \cllamabaseB{34}, and \magicoders{} (\magicoderBLit{}) from \cllamabaseB{7} and \deepseekbaseB{6.7}.
See Appendix~\ref{app:subsec:models} for a complete list of models and estimated cutoff dates. 

\vspace{2pt}
\noindent \textbf{Evaluation Metrics.}
We use the \passmetric{1}~\citep{kulal2019spoc,chen2021evaluating} metric for our evaluations. 
Specifically, we generate $10$ candidate answers for each problem either using API or using \smalltextsc{vLLM}~\citep{kwon2023efficient}.
We use nucleus sampling with temperature $0.2$ and top\_p $0.95$ and calculate the fraction of programs or answers that are correct. 
For the code generation and self-repair scenarios, we use tests to verify the correctness of the programs. For these scenarios, programs must pass all tests to be considered correct.
For the code execution scenario, we use an execution-based correctness metric between the generated output and the ground truth output. For the test output prediction scenario, we parse the generated response to extract the answer and use equivalence checks for grading as specified in Section~\ref{sec:tasks}.

\subsection{Scenario-specific setup}
\label{subsec:scenario_specific_setup}
The setup for each scenario is presented below.
Note that the base models are only used in the code generation scenario since they do not easily follow the format for the other scenarios.

\vspace{2pt}
\noindent \textbf{Code Generation.} 
For the instruction-tuned models, we use a zero-shot prompt and follow the approach of \citet{hendrycks2021measuring} by adding appropriate instructions to generate solutions in either functional or stdin format.
For the base models, we use a constant one-shot example, with a separate example provided for problems that accept stdin input and for problems that accept functional output.
Section~\ref{appendix:codegen-prompts} shows the high-level zero-shot prompt used.

\vspace{2pt}
\noindent \textbf{Self Repair.} 
Similar to prior work~\cite{olausson2023demystifying}, we use the programs generated during the code generation scenario 
along with the corresponding error feedback to build the zero-shot prompt for the self-repair scenario.
The type of error feedback includes syntax errors, runtime errors, wrong answers, and time-limit errors, as applicable.
Section~\ref{appendix:repair-prompts} provides the pseudo-code for computing the error feedback and the corresponding prompt.

\vspace{2pt}
\noindent \textbf{Code Execution.} 
We use few-shot prompts for the code execution scenario, both with and without chain-of-thought prompting (\chainot{}). 
Particularly, we use a 2-shot prompt without \chainot{} and a 1-shot prompt with \chainot{} with manually detailed steps.
The prompts are detailed in Section~\ref{appendix:code-execution-prompts}.

\vspace{1pt}
\noindent \textbf{Test Output Prediction.} 
We use a zero-shot prompt that queries the model to complete assertions, given the problem, function signature, and test input. 
We provide the prompt in Section~\ref{appendix:test-pred-prompts}.

\section{Results}
\label{sec:results}
We first describe how \livecodebench{} helps detect and avoid benchmark contamination in Section~\ref{subsec:contamination}. 
Next, we present the findings from our evaluations on \livecodebench{} in Section~\ref{subsec:model_comparisions}.


\subsection{Avoiding Contamination}
\label{subsec:contamination}
A distinguishing aspect of our benchmark is the ability to evaluate models on problems released over different time windows.
This allows us to measure the model performance on problems released after the cutoff date, thereby giving a performance estimate on \textit{unseen} problems.

\vspace{1pt}
\noindent \textbf{Contamination in \deepseek{} and \gptfouro{}.}
\livecodebench{} comprises problems released since May $2023$. 
However, \deepseek{} models were released in Sep $2023$ and might have already been trained on some of the problems in our benchmark. 
Similarly, \openai{} notes \gptfouro{} cutoff date in November. 
We can measure the performance of the models on the benchmark using problems released after the cutoff date, thereby estimating the performance of the model on previously unseen problems.
Figure~\ref{fig:contamination_main} shows the performance of these models on \livecodebench{} code generation and test output prediction scenario on \leetcode{} problems released in different months from May $2023$ and Feb $2024$. 
We notice a stark drop in the performance of \deepseekcodeB{33} model after Aug. $2023$ (right before its release date), which suggests 
that the earlier problems might indeed be contaminated.
This trend is consistent across other \livecodebench{} scenarios like repair and code execution, as depicted in Figure~\ref{fig:contamination_all_tasks_line}.
Concurrently, \citet{deepseek-coder} (Section 4.1, last paragraph) also acknowledge the possibility of \leetcode{} contamination, noting that ``\textit{models achieved higher scores in the LeetCode Contest held in July and August}.
Similarly, performance of the \gptfouro{} model drops on problems released since November (its official cutoff date).

Interestingly, we find that this drop in performance primarily occurs for the \leetcode{} problems only and that the model performance is relatively smooth across the months for problems from other platforms. Figure~\ref{fig:atcoder_contamination} shows a relatively stable performance for all models on \atcoder{} problems released over different periods, with the possible exception of May and June. 





\vspace{1pt}
\noindent \textbf{Performances of other models.}
We study performance variations in other models released recently.
Particularly, \gptfourturbo{}, \geminipro{}, \mistrallarge{}, and \claudethrees{} models were released in November $2023$, December $2023$, February $2024$, and March $2024$ respectively.
Note that \gptfourturbo{} ($1106$-preview variant) and \claudethrees{} have cutoff dates April $2023$ and August $2023$ respectively.
Irrespective of the release or cutoff dates, we do not find any drastic performance variations across the months, as shown in Figure~\ref{app:fig:contamination_all_models}, particularly compared to the \deepseek{} models.
Interestingly, we find that even the \deepseekbaseB{33} model also suffers from contamination dropping from \passmetric{1} $\smallsim 60$ in May problems to \passmetric{1} $\smallsim 0$ in September \leetcode{} problems. 
This also suggests the likely inclusion of competition problems in the pretraining of the \deepseek{} models, thereby affecting all instruction models trained from it.
Finally, \codestral{} achieves \passmetric{1} $36.5$ on problems released between May'23 and Jan'24 and \passmetric{1} $28.3$ on problems since Feb'24.

\begin{figure*}[t]
    \centering
      \subfloat{
          \includegraphics[width=0.48\linewidth]{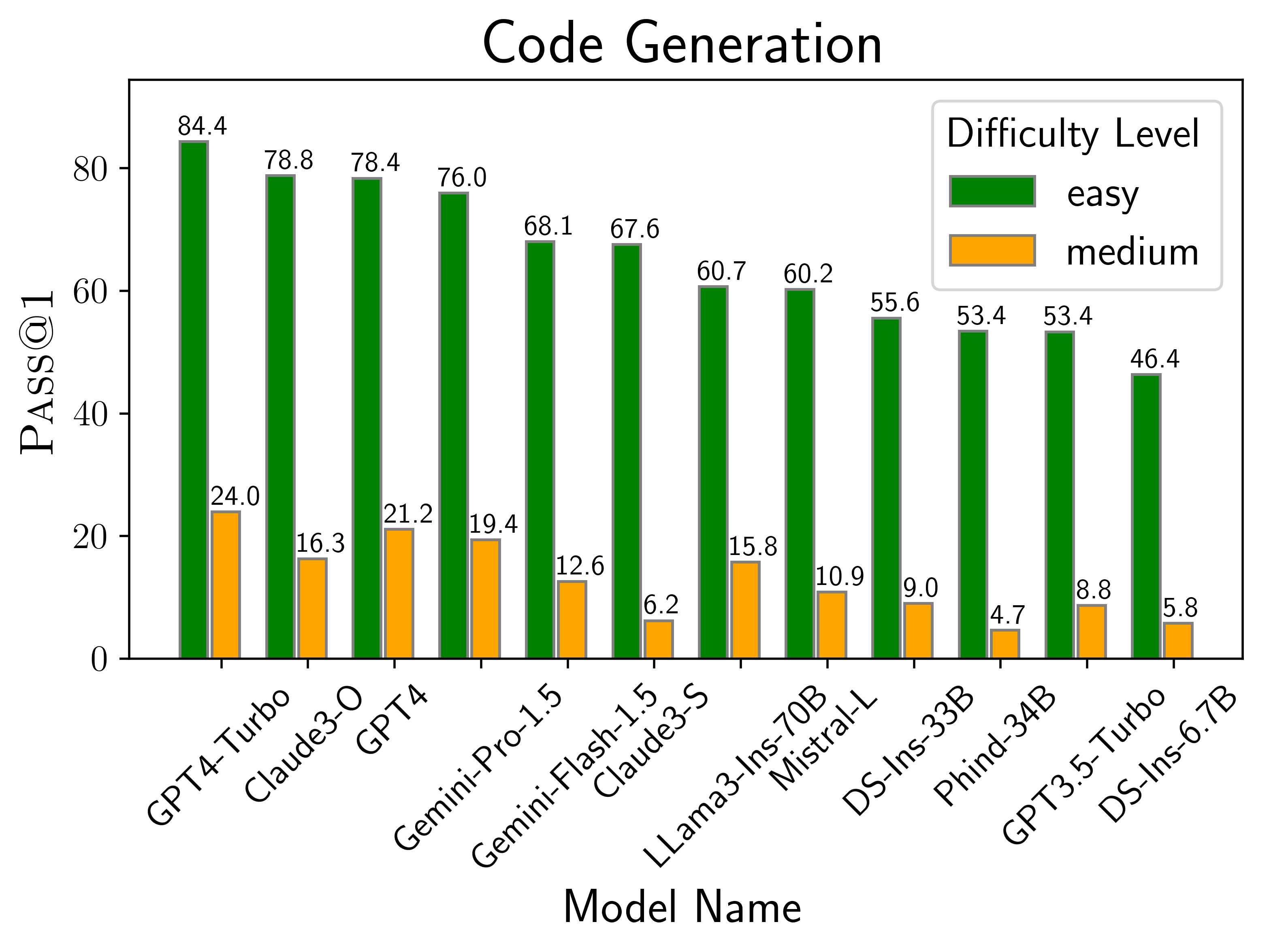}}
      ~
      \subfloat{
          \includegraphics[width=0.48\linewidth]{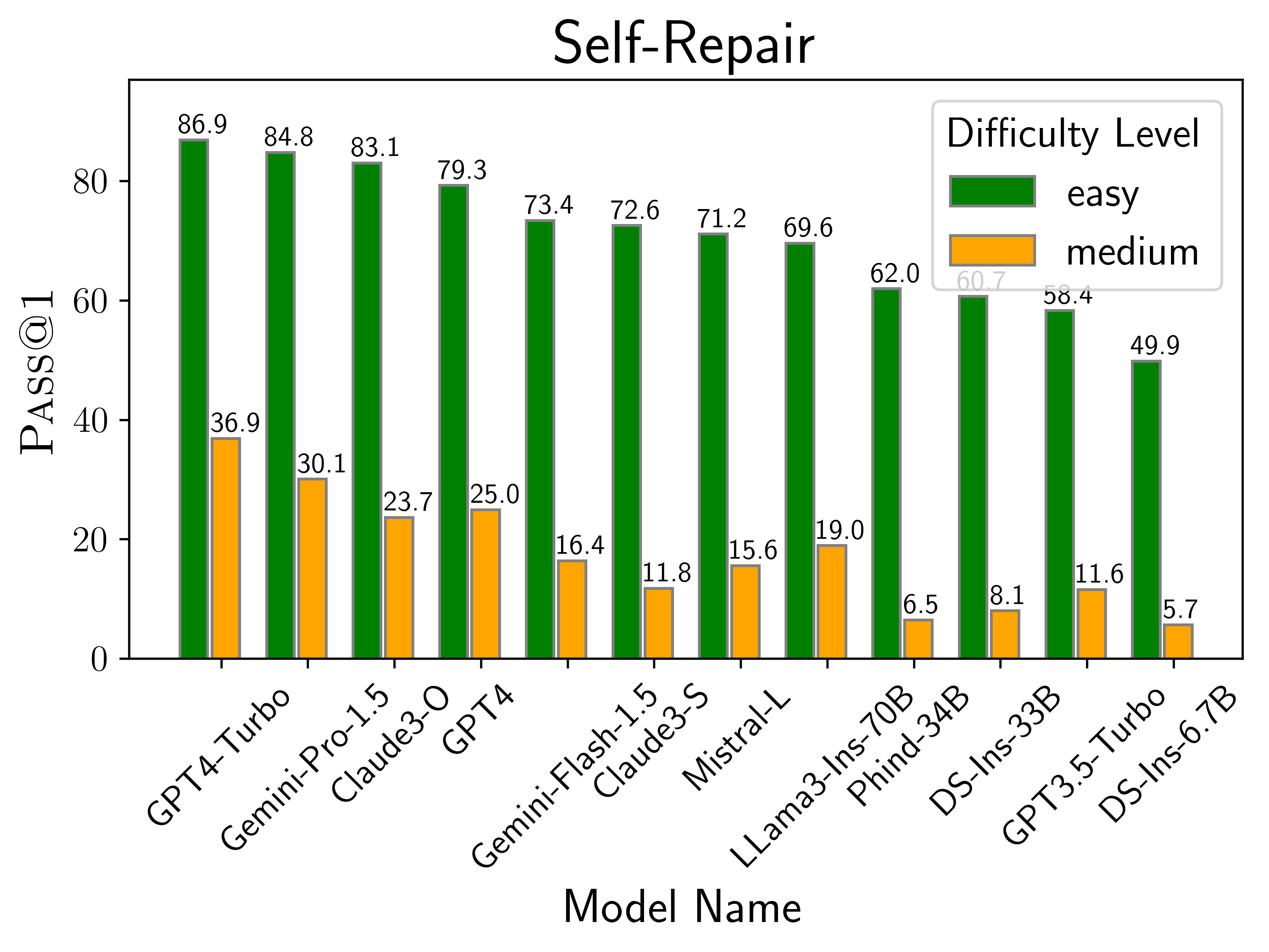}}
      
      \subfloat{
          \includegraphics[width=0.48\linewidth]{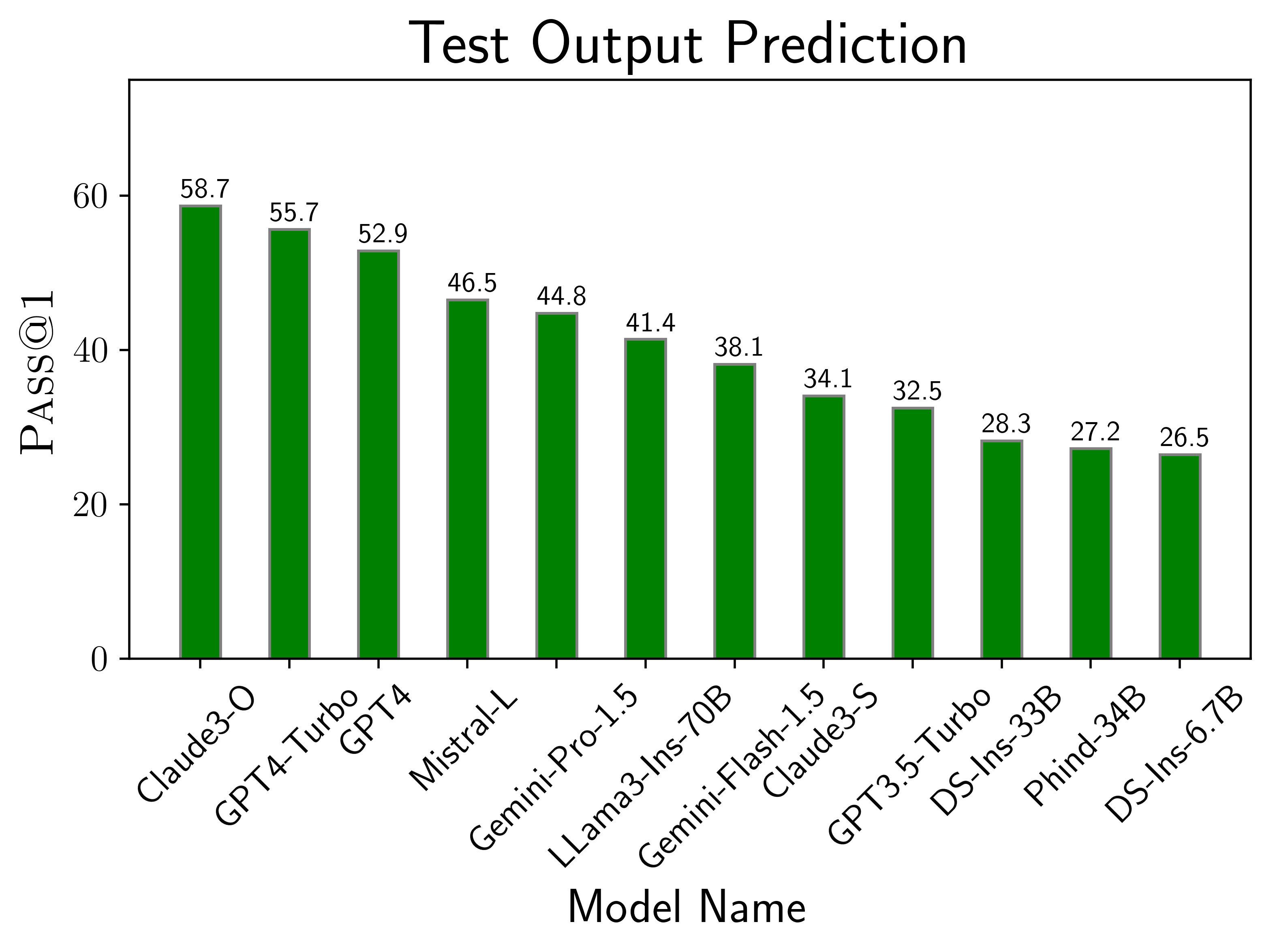}}
      ~
      \subfloat{
        \includegraphics[width=0.48\linewidth]{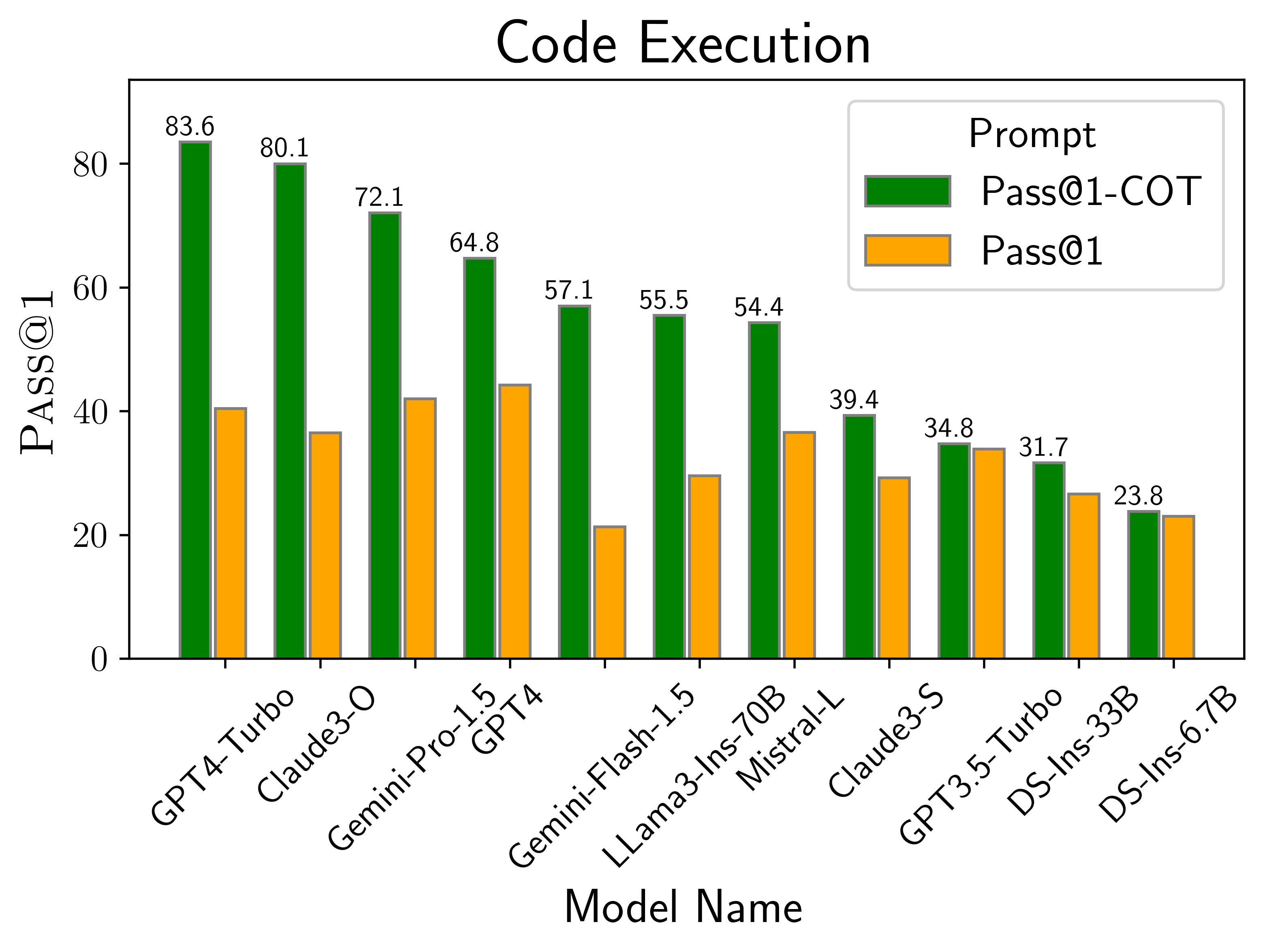}}
  
  \vspace{-2pt}
  \caption{
    Model performances across the four scenarios available in \livecodebench{} (filtering on the time-window post September).
    The top-left and top-right plots depict \passmetric{1} of models on easy and medium splits across the code generation and self-repair scenarios respectively (results on hard subset deferred to the Appendix).
    The bottom-left and bottom-right plots depict \passmetric{1} of models across the test output prediction and code execution scenarios respectively.
  }
  \vspace{-10pt}
  
  \label{fig:results_plots}
  \end{figure*}

\subsection{Performance and Model Comparisons}
\label{subsec:model_comparisions}
We evaluate $34$ instruction-tuned models (and $18$ base models used in the code generation scenario) on \livecodebench{}.
These models range from closed access to open access with their various fine-tuned variants.
To overcome contamination issues in \deepseek{} models, we only consider problems released since Sep $2023$ for all evaluations below.
Figure~\ref{fig:results_plots} shows the performance of a subset of models across the four scenarios. 
We highlight our key findings below.

\vspace{1.5pt}
\noindent \textbf{Holistic Evaluations.}
We have evaluated the models across the four scenarios currently available in \livecodebench{}.
Figure~\ref{fig:all_tasks_radial} displays the performance of models on all scenarios along the axes of the polar chart.
First, we observe that the relative order of models remains mostly consistent across the scenarios.
This is also supported by high correlations between \passmetric{1} metric across the scenarios -- 
over $0.88$ across all pairs as shown in Figure~\ref{app:fig:correlation_matrix}.
Interestingly, the correlations are larger for related tasks, $0.98$ for generation and self-repair, and $0.96$ for test output prediction and code execution. 
This correlation drops to $0.89$ for generation and execution scenarios.

However, despite the strong correlation, the relative differences in performance do vary across the scenarios.
For example, \gptfourturbo{} further gains performance gap over \gptfour{} in the self-repair scenario after already leading in the code generation scenario.
Similarly, \claudeopus{} and \mistrallarge{} perform well in tasks involving \chainot{}, particularly in the code execution and test output prediction scenarios.
For instance, \claudeopus{} even outperforms \gptfourturbo{} in the test output prediction scenario.
Similarly, \mistrallarge{} outperforms \claudesonnet{} in both scenarios after trailing behind in code generation and repair scenarios.
These differences highlight the need for holistic evaluations beyond measuring code generation capabilities.


\begin{figure}[!t]
    \centering
    \includegraphics[width=\textwidth]{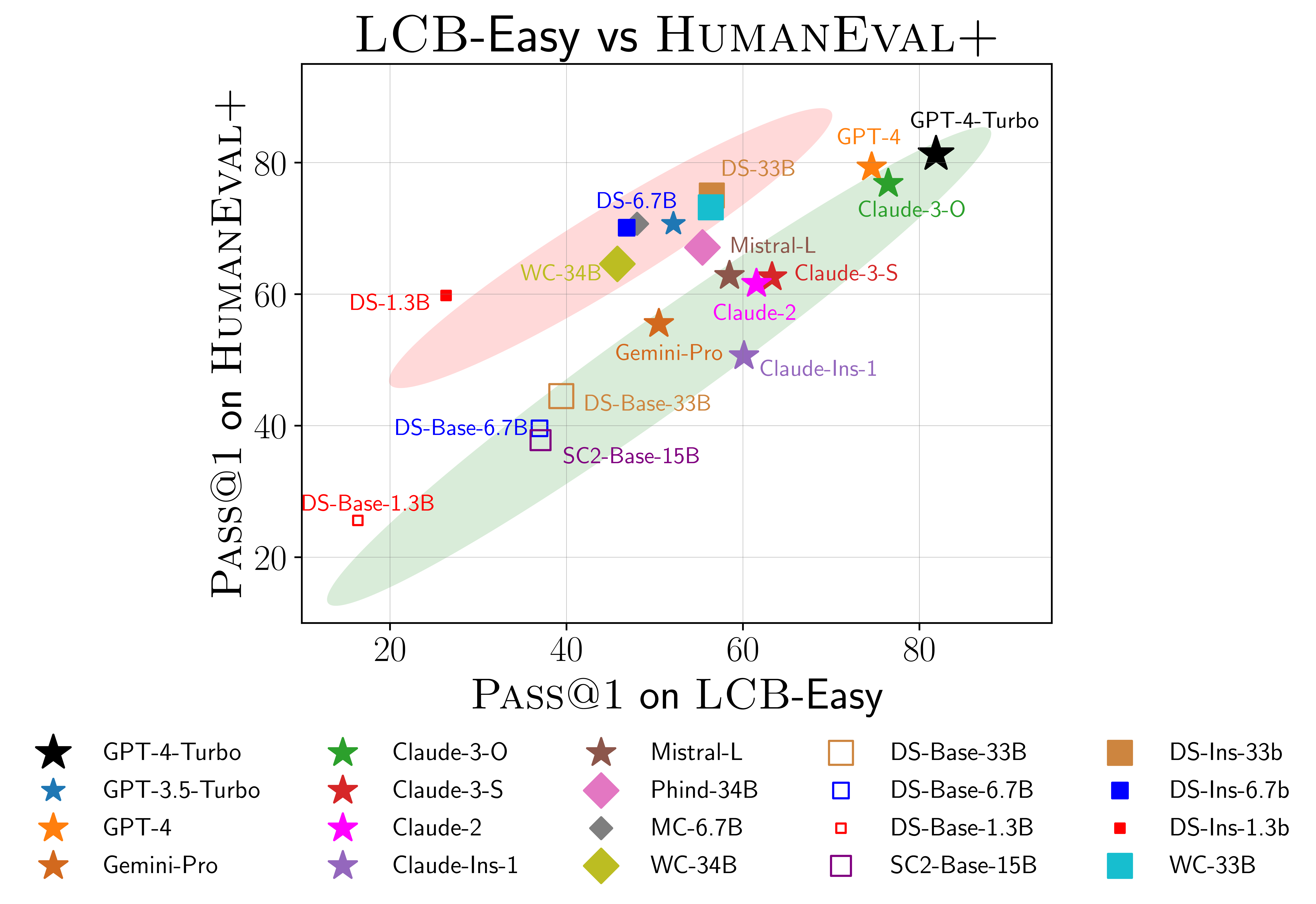}
  \vspace{-3pt}
    \caption{
        Scatter plot comparing \passmetric{1} of models on \humanevalplus{} versus \passmetric{1} on the easy subset of \livecodebench{} code generation scenario.
        Star markers denote the closed-access models while other markers denote different open model families. 
        We find that the models are separated into two groups -- the green-shaded region where performances on the two datasets are \textit{aligned} and the red-shaded region where models perform well on \humanevalplus{} but perform poorly on \livecodebench{}.
        This indicates potential overfitting on \humanevalplus{} and primarily occurs in the fine-tuned variants of open-access models. 
        For example, \deepseekcodeB{1.3} which achieves \passmetric{1} of $60$ and $26$ on \humanevalplus{} and \lcb{}-Easy subset.
        Thus, while it ranks above \commandrplus{} on \humanevalplus{}, it performs significantly worse on the \lcb{}.
        Similarly, \deepseekcodeB{6.7} and \codeqwen{} outperform \claudesonnet{} on \humanevalplus{} but are $>20$ points behind on \lcb{}-Easy. 
    }
    \vspace{-5pt}
    \label{fig:he_vs_lcb}
\end{figure}

\vspace{1pt}
\noindent \textbf{Comparison to \humaneval{}.}
Next, we compare how code generation performance metrics translate between \livecodebench{} and \humaneval{}, 
the primary benchmark used for evaluating coding capabilities.
Note that we use \humanevalplus{} version of \humaneval{} problems as it is more reliable with more exhaustive test cases.
Figure~\ref{fig:he_vs_lcb} shows a scatter plot of \passmetric{1} on \humanevalplus{} versus \lcb{}-Easy code generation scenario. 
We find only a moderate correlation of $0.72$, with much larger performance variations on \lcb{}-Easy.

Additionally, we observe that the models cluster into two groups, shaded in red and green.
The models in the green-shaded region lie close to the $x=y$ line, indicating that they perform similarly on both benchmarks.
On the other hand, the models shaded in red lie in the top-left region of the graph, 
indicating that they perform well only on \humanevalplus{} but not as well on \livecodebench{}.
Interestingly, the green-shaded cluster contains base models or closed-access models, 
while the red-shaded cluster primarily comprises fine-tuned variants of open-access models.
The well-separated clusters suggest that many models that perform well on \humaneval{} might be overfitting on the benchmark, and their performances do not translate well to problems from other domains or difficulty levels like those present in \livecodebench{}. 

 
Indeed, \humaneval{} is an easier benchmark with small and isolated programming problems and thus easier to overfit on.
In contrast, \livecodebench{} problems are sourced from reputable coding platforms offering more challenging problems with higher diversity and difficulty levels.
This potential overfitting is particularly exemplified by \deepseekcodeB{1.3} which achieves $59.8$\% \passmetric{1} on \humanevalplus{} but only $26.3$\% on \lcb{}-Easy.
Thus, while it boasts better performance compared with \geminipro{} and \claudeinstantone{} on \humanevalplus{}, it performs considerably worse on \lcb{}-Easy.
Similarly, \codeqwen{}, \deepseekcodeB{6.7}, and \magicoder{6.7} perform better than \mistrallarge{}, \claudetwo{}, and \claudesonnet{} on \humanevalplus{} but are considerably worse on \lcb{}-Easy.


\vspace{1pt}
\noindent \textbf{Highlighting the gap between SoTA and open models.} 
One distinct observation from our evaluations is the large gap between SoTA models and open models across all scenarios.
Particularly, \gptfourturbo{}, \gptfour{}, \geminiproonefive{} and \claudeopus{} lead across the benchmarks with wide performance margins over other models.
This distinguishes \livecodebench{} from prior benchmarks (like \humaneval{}) where various open models have achieved similar or better performance. 
For example, \deepseekcodeB{33} is merely $4.3$ point behind \gptfourturbo{} on \humanevalplus{} but 
$16.2$ points ($69$\%) on \lcb{} code generation scenario.
This gap either holds or sometimes even amplifies across other scenarios.
For instance, consider test output prediction and code execution (with \chainot{}) where \gptfourturbo{} leads the \deepseekcodeB{33} model by $96\%$ and $134\%$ respectively!

We qualitatively analyze code samples generated by the leading model, \gptfourturbo{}, and find that it generates more readable code.
Specifically, the code consists of more inline natural language comments that \textit{reason} or \textit{plan} before producing the code.
We verify this quantitatively and find \gptfourturbo{} generated uses $19.5\times$ more comment tokens compared to \gptfour{}.

\vspace{1pt}
\noindent \textbf{Comparing Base Models.}
We use four families of base models -- \llamabase{}, \deepseek{}, \cllama{}, and \starcodertwo{} and compare them on the code generation scenario.
A one-shot prompt is used for all models to avoid any formatting and answer extraction issues.
We find \llamabase{} and \deepseekcode{} models are significantly better than both \cllama{} and \starcodertwo{} base models with 
a \deepseekbaseB{6.7} model even outperforming both \cllamabaseB{34} and \sctwobaseB{15} models.
Next, we observe that \sctwobaseB{15} also outperforms the \cllamabaseB{34} model (similiar to findings in ~\cite{starcoder2}).
Note that some \livecodebench{} specific differences can potentially be attributed to data curation approaches.
For instance, \starcodertwo{} models (and potentially \deepseeks{} as discussed in Section~\ref{subsec:contamination}) use competition problems in the pre-training corpus.

\vspace{1pt}
\noindent \textbf{Role of Post Training.}
We find that post-training improves performance on both \humanevalplus{} and \livecodebench{} for the code generation scenario.
Particularly, \llamaB{70}, \deepseekcodeB{33} and \phind{} achieve $28.3$, $23.6$, and $21$ \passmetric{1} on \lcb{} 
improving over their base models by $8.2$, $7.3$ and $9.5$ points respectively.
Similar gains are observed in previous benchmarks (like \humanevalplus{}) as well.
This highlights the importance of good post-training datasets  for building strong \llms{}. 

At the same time, we note that the base models have \textit{aligned} performances on \lcb{} code generation and \humanevalplus{} benchmarks and lie within or close to the green shaded region in Figure~\ref{fig:he_vs_lcb}.
However, the fine-tuned open models exhibit a larger performance gap, with much better performances on \humanevalplus{}.
On the other hand, the closed-access models are still aligned across both benchmarks.
This suggests that the fine-tuning data for open models might not be as diverse as that for closed models, leading to a lack of generalization to different kinds of problems.

\vspace{1pt}
\noindent \textbf{Comparing open-access instruction-tuned models.}
Here, we compare various fine-tuned variants of the \llamabase{}, \deepseek{} and \cllama{} base models across different model sizes.
We find that fine-tuned \llamabase{} and \deepseek{} models lead in performance, followed by \phind{} and \cllama{} models across most scenarios.
Broadly, we find that model performances correlate with model sizes. For example, \phind{} model outperforms the $6.7$\smalltextsc{B} models across all scenarios.


\vspace{1pt}
\noindent \textbf{Comparing Closed Models.}
We evaluate a range of closed (API access) models ranging from different model families 
like \gpts{}, \claudes{}, \gemini{}, and \mistral{}.
We find the \gptfourturbo{} and \claudeopus{} rank at the top across all scenarios followed by \mistrallarge{} and \claudesonnet{} models. 
Finally, \geminipro{} and \gptthreefiveturbo{} lie on the lower end of the models.
The relative differences between the models vary across the scenarios. 
For example, \gptfourturbo{} demonstrates remarkable improvement from self-repair ($24.5$\% to $36.9$\% on the \lcb{}-Medium problems) while \geminipro{} only improves from $8.5$\% to $9.4$\%.
Similarly, as identified above, \claudeopus{} and \mistrallarge{} perform considerably better on test output prediction and code execution scenarios. 

\vspace{1pt}
\noindent \textbf{Open-Access vs Closed-Access Models.}
In general, closed (API) access model families generally outperform the open access models.
The gap is only closed by three models, namely \llamaB{70}, \mixtral{}, and \deepseekcodeB{33}
which reach the performance levels of the closed models.
For instance, in the code generation scenario (Figure~\ref{fig:all_tasks_radial} right), 
these models reach close to or even outperform closed access models like \geminipro{}, \gptthreefiveturbo{}, and \claudesonnet{}.
The performances vary across scenarios with the closed-access models performing better in test output prediction and code execution scenarios. 
Overall, our findings confirm that a combination of strong base models and high-quality post-training datasets is a viable recipe for good code \llms{}. 




\section{Related Work}
\label{sec:related-work}
\subsection{Code Generation}
\noindent \textbf{Language Models for Code Generation.} 
Starting with Codex~\citep{chen2021evaluating}, there are over a dozen code \llms{}. These include CodeT5~\citep{wang2021codet5, wang2023codet5+}, CodeGen~\citep{nijkamp2022codegen}, SantaCoder~\citep{allal2023santacoder}, StarCoder~\citep{li2023starcoder}, AlphaCode~\citep{li2022competition}, InCoder~\citep{fried2022incoder}, and CodeGeeX~\citep{zheng2023codegeex}. 
As of May 2024, \llamabase{} and \deepseek{}~\citep{bi2024deepseek}, \starcoder{}~\cite{starcoder2,li2023starcoder} and \cllama{}~\citep{roziere2023code} are the most popular open models. 
Many downstream models resulted from fine-tuning them on synthetically generated data, such as \wizardlong{} \citep{luo2023wizardcoder}, \magicoders{} \citep{wei2023magicoder}, and \phind{}.

\vspace{1pt}
\noindent \textbf{Code Generation Benchmarks.} 
Many benchmarks have been proposed to compare and evaluate these models. 
These primarily focus on natural language to Python code generation: 
\humaneval{} \citep{chen2021evaluating}, 
\humanevalplus{} \citep{liu2023your}, 
\apps{} \citep{hendrycks2021measuring}, 
\contests{} \citep{li2022competition}, 
\mbpp{} \citep{austin2021program}, 
L2CEval \citep{ni2023l2ceval}. 
Their variants have been proposed to cover more languages, \citep{wang2022recode, zheng2023codegeex, cassano2022multipl, athiwaratkun2022multi}.
Many benchmarks have focused on code generation in APIs. Benchmarks like DS-1000 \citep{lai2023ds}, ARCADE \citep{yin2022natural}, NumpyEval \citep{zhang2023toolcoder}, and PandasEval \citep{jain2022jigsaw} focus on data science APIs. 
Other benchmarks measure using broader APIs or 
general software engineering tasks, such as JuICe~\citep{agashe2019juice}, APIBench~\citep{patil2023gorilla}, RepoBench~\citep{liu2023repobench}, ODEX~\citep{wang2022execution}, SWE-Bench~\citep{jimenez2023swe}, GoogleCodeRepo~\citep{shrivastava2023repository}, RepoEval~\citep{zhang2023repocoder}, and Cocomic-Data~\citep{ding2022cocomic}.~

A few benchmarks specifically measure competitive programming, such as 
APPS \citep{hendrycks2021measuring}, 
CodeContests \citep{li2022competition}, 
CodeScope \citep{yan2023codescope},
xCodeEval \citep{khan2023xcodeeval},
and LeetCode-Hard \citep{shinn2023reflexion}, and TACO \citep{li2023taco}. 
Methods such as AlphaCode \citep{li2022competition}, AlphaCode 2\citep{team2023gemini}, ALGO \citep{zhang2023algo}, Parsel \citep{zelikman2022parsel}, code cleaning \citep{jain2023llm}, code explainations \citep{li2023explaining}, analogical reasoning \citep{yasunaga2023large}, and AlphaCodium \citep{ridnik2024code} have been pushing the boundaries of what is possible with \llms{} in this domain. The biggest differentiating factor between \livecodebench{} and these benchmarks is that our benchmark is \textbf{continuously updated}, \textbf{problem curation with balanced difficulty}, \textbf{higher tests and problem quality}, and contains \textbf{more scenarios} such as code repair, code execution, and test output prediction capturing more facets for building agentic coding systems.

\subsection{Holistic Tasks}
\livecodebench{} considers self-repair, test output prediction, and code execution as additional scenarios. Below we note pertinent related work for these domains. 


\vspace{1pt}
\noindent \textbf{Code Repair.} 
\citep{chen2023teaching, olausson2023demystifying, madaan2023self, peng2023check, zhang2023self} have investigated self-repair for existing code \llm{} benchmarks. Particularly, these methods use error feedback for models to improve inspiring our code repair scenario.

\vspace{1pt}
\noindent \textbf{Code Execution.} 
Code execution was first studied in \citet{austin2021program, nye2021show}
\livecodebench{}'s execution scenario is particularly inspired by CRUXEval \citep{gu2024cruxeval}, a recent benchmark measuring the reasoning and execution abilities of code \llms{}. 
We differ from CRUXEval in that our benchmark is live, and our functions are more complex and human-produced (unlike Code Llama generations in CRUXEval).

\vspace{1pt}
\noindent \textbf{Test Generation.} 
Test generation using \llms{} has been explored in \citep{yuan2023no, 10329992, tufano2022methods2test, watson2020learning}. 
Furthermore, \citet{chen2022codet} demonstrated that \llms{} can assist in generating test case inputs/outputs for competitive programming problems, thereby improving the accuracy of the generated code, thus inspiring our test generation scenario.
However, \livecodebench{}'s test generation scenario is unique in that it decouples the test inputs and outputs allowing more proper evaluations.

Finally, some works have additionally studied other tasks and scenarios like type prediction \citep{mir2022type4py, wei2023typet5, malik2019nl2type}, code summarization \citep{leclair2019neural, iyer2016summarizing, barone2017parallel, hasan2021codesc, alon2018code2seq}, code security \citep{liguori2022can, pearce2022asleep, tony2023llmseceval}, etc. 

\subsection{Contamination}
Data contamination and test-case leakage have received considerable attention ~\cite{yonatan2024proving,golchin2023time,weller2023according,roberts2024to} as \llms{} might be getting trained on benchmarks. 
~\citet{sainz2023chatgpt} demonstrated contamination by simply prompting the model to highlight its contamination. 
Some detection methods have also been built to avoid these cases~\citep{shi2023detecting,zhou2023don}. For code, \citet{riddell2024quantifying} use edit distance and AST-based semantic-similarity to detect contamination. 

\section{Limitations}
\vspace{1pt}
\noindent \textbf{Benchmark Size.}
\livecodebench{} code generation scenario currently hosts over $400$ instances from problems released between May and February.
To account for contamination in \deepseek{}, we only perform evaluations on problems released after the model cutoff date.
This leads to only $349$ problems used in our final evaluations which might add noise due to problem set samples.
We currently estimate $1-1.5\%$ performance variations in \livecodebench{} code generation due to this issue (measured by bootstrapping $349$ sized problem sets from the $511$ sized dataset).
Other scenarios, i.e. self-repair, code execution, and test output prediction comprise $349$, $188$, and $254$ problems would have similar performance variations. 
We thus recommend exercising proper judgement when comparing models with small performance differences.
Note that \humaneval{} has $164$ problems and would also struggle with similar issues.

This issue is also exacerbated for newer models, with more recent cutoff dates, as they might only have access to a smaller evaluation set.
We propose two solutions addressing this issue as we evolve \livecodebench{}.
First, we will use other competition platforms for problem collection, allowing larger number of recent problems to be added to the benchmark.
In addition, we also hope supplement this with an unreleased private test set constructed specifically for model evaluation. 
These problems will use a similar flavor to current problems and will be used when models are submitted for evaluation to the \livecodebench{} platform.
This would reduce the reliance on public accessible problems and provide a more robust evaluation of the models while providing community public access to similar problems, similar to strategies employed by popular platforms like \smalltextsc{Kaggle}.

\vspace{1pt}
\noindent \textbf{Focus on \python{}.} 
\livecodebench{} currently only focuses on \python{} which might not provide enough signal about model capabilities in other languages. 
However, since we collected problem statements and serialized tests, adding new programming languages would be straightforward once appropriate evaluation engines are used.

\vspace{1pt}
\noindent \textbf{Robustness to Prompts.}
Recent works have identified huge performance variances that can be caused due to insufficient prompt.
Here, we either do not tune prompts across models or make minor adjustments based on the system prompts and delimiter tokens.
This can lead to performance variance in our results. 
Our findings and model comparison orders generalize across \livecodebench{} scenarios 
and mostly match the performance trends observed on \humaneval{} making this a less prominient issue.

This issue can be particularly observed open models on the code execution scenario with \chainot{} prompting.
Interestingly, often the open models perform even worse in comparsion to the direct code execution baseline.
Note that we used same prompts for the closed models all of which show noticable improvement from \chainot{}.
While the used prompts might be sub-optimal, this highlights how open-models perform worse against the closed models at performing chain-of-thought.

\vspace{1pt}
\noindent \textbf{Problem Domain.}
Programming is a vast domain and occurs in various forms such as programming puzzles, competition programming, and real-world software development.
Different domains might have individual requirements, constraints, challenges, and difficulty levels.
\livecodebench{} currently focuses on competition problems sourced from three platforms.
This might not be representative of the ``most general'' notion of \llm{} programming capabilities. 
Particularly, real-world usage of \llms{} is drawn upon open-ended and unconstrained problems rasied by users.
We therefore recommend using \livecodebench{} as a starting point for evaluating \llms{} and 
further using domain-specific evaluations to measure and compare \llms{} in specific settings as required.

\section{Conclusion}
\label{sec:conclusion}
In this work, we propose \livecodebench{}, a new benchmark for evaluating \llms{} for code.
Our benchmark mitigates contamination issues in existing benchmarks by introducing live evaluations and emphasizing scenarios beyond code generation to account for the broader coding abilities of \llms{}.
\livecodebench{} is an extensible framework, that will keep on updating with new problems, scenarios, and models.
Our evaluations reveal novel findings such as contamination detection and potential overfitting on \humaneval{}.
We hope \livecodebench{} with serve to advance understanding of current code \llms{} and also guide future research in this area through our findings.

\section*{Acknowledgements}
This work was supported in part by NSF grants CCF:1900968, CCF:1908870 and by SKY Lab industrial sponsors and affiliates Astronomer, Google, IBM, Intel, Lacework, Microsoft, Mohamed Bin Zayed University of Artificial Intelligence, Nexla, Samsung SDS, Uber, and VMware. 
A. Gu is supported by the NSF Graduate Research Fellowship under Grant No. 2141064. 
A. Solar-Lezama is supported by the NSF and Intel Corporation through NSF Grant CCF:2217064.
Any opinions, findings, conclusions, or recommendations in this paper are solely those of the authors and do not necessarily reflect the position of the sponsors. 

\noindent Finally, we thank  Manish Shetty, Wei-Lin Chiang, Jierui Li, Horace He, Federico Cassano, Pengcheng Yin, and Aman Madaan for helpful feedback at various stages of the work. 

\bibliography{bibliography}
\bibliographystyle{acl_natbib}

\newpage
\appendix

\section{Dataset}

\subsection{License}
Similar to \citet{hendrycks2021measuring}, we scrape only the problem statements, ground-truth solutions, and test cases from competition websites -- \leetcode{}, \atcoder{}, and \codeforces{}. 
Further, we only scrape publicly visible portions of websites, avoiding any data collection that might be pay-walled or require login or interaction with the website.
Following, \citet{hendrycks2021measuring} we abide by Fair Use §107: “the fair
use of a copyrighted work, including such use by ... scholarship, or research, is not an infringement
of copyright”, where fair use is determined by “the purpose and character of the use, including
whether such use is of a commercial nature or is for nonprofit educational purposes”, “the amount
and substantiality of the portion used in relation to the copyrighted work as a whole”, and “the effect
of the use upon the potential market for or value of the copyrighted work.” 
Finally, we use the collected problems for academic purposes only and in addition, do not train on the collected problems.


\subsection{Generator Based Test Generation}
\label{appendix:input-generators}
We use \gptfourturbo{} to construct input generators. 
The following prompts (Figures~\ref{fig:rand_inp_gen} and \ref{fig:adv_inp_gen}) provide one-shot prompt templates used for synthesizing random and adversarial input generators.
These generators define a function returns the arguments sampled in some distribution.
These generators are then executed to construct inputs which validated on the collected correct programs.
We use separate generators for random and adversarial setting since often times programming problems have corner cases which might not be captured by randomly sampling over the inputs. We build $2$ random input generators, 4 adversarial input generators and check if the sampled inputs work for the correct programs. Finally, the number of collected inputs is thresholded to $100$ for efficient grading (using random selection). We find that our generators can already function well but future work can study the design space of constructing such generators. 

Note that for \codeforces{}, we construct the generators in semi-autonomous manner since only $9$ problems were used.

\begin{figure*}[h!]
\centering
\begin{tcolorbox}[enhanced,size=small,colback=black!5!white,colframe=RoyalBlue,flip title={interior hidden},title={Prompt for random input generation}]
\begin{lstlisting}[label={lst:execution-example-filtered-1}, captionpos=b, breaklines=true, language=Python]
You are an expert python competitive programmar and your goal is to construct input-generators for testing programming contest problems. You will write relevant generators and finally construct `construct_inputs` function that returns a list of diverse inputs sampled from the generator. Remember to strictly follow the instructions and constraints present in the problem statement

<USER>
{EXAMPLE_PROBLEM}

Construct a random input generator.
</USER>

<ASSISTANT>
import numpy as np
def random_input_generator(weight_min, weight_max, size_min, size_max):
    weights_size = np.random.randint(size_min, size_max+1)
    weights = np.random.randint(weight_min, weight_max, size=weights_size).tolist()
    k = np.random.randint(1, len(weights)+1)
    return weights, k

def construct_inputs():
    inputs_list = []
    ## small inputs
    for i in range(15):
        inputs_list.append(random_input_generator(1, 10**3, 1, 10))
    ## medium inputs
    for i in range(15):
        inputs_list.append(random_input_generator(1, 10**6, 1, 10**3))
    ## large inputs
    for i in range(15):
        inputs_list.append(random_input_generator(1, 10**9, 1, 10**5))
    return inputs_list
</ASSISTANT>

<USER>
{PROBLEM}

Construct a random input generator. Use the format used in the above example by returning a single function that builds diverse inputs named `construct_inputs`
</USER>
\end{lstlisting}
\end{tcolorbox}
\caption{Random Input Generator Prompt}
\label{fig:rand_inp_gen}
\end{figure*}

\begin{figure*}[h!]
\centering
\begin{tcolorbox}[enhanced,size=small,colback=black!5!white,colframe=RoyalBlue,flip title={interior hidden},title={Prompt for adversarial input generation}]
\begin{lstlisting}[label={lst:execution-example-filtered-1}, captionpos=b, breaklines=true, language=Python]
You are an expert python competitive programmar and your goal is to construct input-generators for testing programming contest problems. You will write relevant generators and finally construct `construct_inputs` function that returns a list of diverse inputs sampled from the generator. Remember to strictly follow the instructions and constraints present in the problem statement

<USER>
{EXAMPLE_PROBLEM}

Construct an adversarial input generator.
</USER>

<ASSISTANT>
import numpy as np
## case 1 - alternating large and small weights
def generate_adversarial_inputs_1(weight_size, max_weight, k):
    weights = [1 if i%2==0 else max_weight for i in range(weight_size)]
    return weights, k

## case 2 - equal_weights
def adversarial_input_generator_2(weight_size, max_weight, k):
    weights = [max_weight for _ in range(weight_size)]
    return weights, k

# Case 3 - Large weights at the ends
def adversarial_input_generator_3(weight_size, max_weight, k):
    pass ## truncatede
def construct_inputs():
    inputs_list = []

    weight_sizes = [10, 1000, 100000]
    max_weights = [10**3, 10**6, 10**9]

    for weight_size in weight_sizes:
        for max_weight in max_weights:
            ks = [1, 2, 5, weight_size//2, weight_size-1, weight_size]
            for k in ks:
                inputs_list.append(generate_adversarial_inputs_1(weight_size, max_weight, k))
                # truncated 
    return inputs_list        
</ASSISTANT>

<USER>
{PROBLEM}

Construct an adversarial input generator. Use the format used in the above example by returning a single function that builds diverse inputs named `construct_inputs`
</USER>
\end{lstlisting}
\end{tcolorbox}
\caption{Adversarial Input Generator Prompt}
\label{fig:adv_inp_gen}
\end{figure*}

\subsection{Code Execution} \label{appendix:dataset-execution}
The code execution split of LiveCodeBench consists of 479 samples from 85 distinct problems. To encourage diversity in our benchmark while keeping our benchmark small and usable, we place a limit of six samples for each given problem. These sample programs and corresponding test cases are chosen uniformly at random from all those passing the filter.

\textbf{Filtering Criteria}: The specific filtering criteria are as follows:
\begin{itemize}
\item Compile time: length of code is between 100 and 500 characters, no syntax errors, all necessary imports are included

\item Runtime: no floating point operations, true division, exp, other integer operations must have at least one argument $\le 3$, string and list operations must have at least one argument with length $\le 3$, finish running in 2 seconds, ``reasonable'' number of steps (roughly, under 1000 Python bytecode operations).
\end{itemize}

We give two examples of two programs that are filtered out in the Listings below. Our final benchmark consists of 479 samples from 85 problems, but will increase in size due to its live nature.

\begin{figure*}[h!]
\centering
\begin{tcolorbox}[enhanced,size=small,colback=black!5!white,colframe=RoyalBlue,flip title={interior hidden},title={Program filtered because of multiplication}]
\begin{lstlisting}[label={}, captionpos=b, breaklines=true, language=Python]
def check(x, t):
    if x == '':
        return t == 0
    if t < 0:
        return False
    for i in range(len(x)):
        if check(x[:i], t - int(x[i:])):
            return True
    return False

@cache
def punishmentNumber(n: int) -> int:
    if n == 0:
        return 0
    ans = punishmentNumber(n-1)
    if check(str(n * n), n):
        ans += n * n
    return ans
assert punishmentNumber(n = 37) == 1478
\end{lstlisting}
\end{tcolorbox}
\end{figure*}

\begin{figure*}[h!]
\centering
\begin{tcolorbox}[enhanced,size=small,colback=black!5!white,colframe=RoyalBlue,flip title={interior hidden},title={Program filtered because of control flow}]
\begin{lstlisting}[label={lst:execution-example-filtered-1}, captionpos=b, breaklines=true, language=Python]
dp = [True for _ in range(int(1e6 + 5))]
MAXN = int(1e6 + 5)
p = []
dp[0] = False
dp[1] = False
for i in range(2, MAXN):
    if not dp[i]: continue
    p.append(i)
    for j in range(2 * i, MAXN, i):
        dp[j] = False
def findPrimePairs(n: int) -> List[List[int]]:
    res = []
    for i in range(1, n):
        if n % 2 == 1 and i > n//2: break
        if n % 2 == 0 and i > n//2: break
        if dp[i] and dp[n - i]:
            res.append([i, n - i])
    return res
assert findPrimePairs(n = 2) == []
\end{lstlisting}
\end{tcolorbox}
\end{figure*}

\begin{figure}[ht!]
    \centering
    \includegraphics[width=0.48\textwidth]{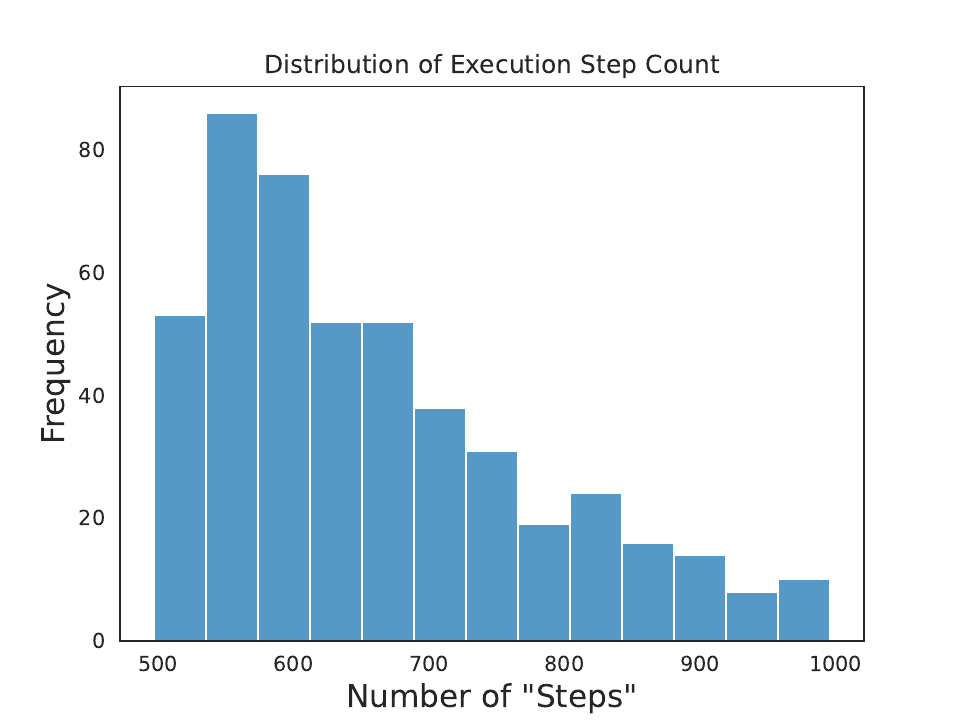}
    \includegraphics[width=0.48\textwidth]{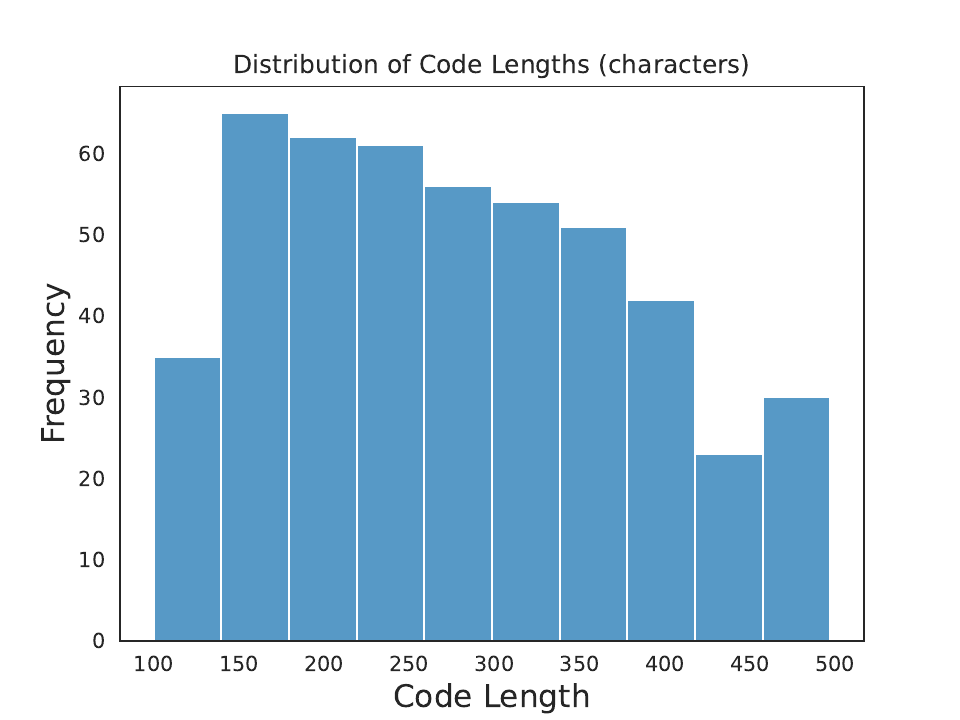}
    \caption{Distribution of code lengths and number of execution steps}
    \label{fig:benchmark-filtering-size-selection}
\end{figure}

\textbf{Dataset Statistics}: As mentioned, we filter for codes between 100 and 500 characters, as well as below 1000 steps. The statistics for programs in our dataset are shown in Fig. \ref{fig:benchmark-filtering-size-selection}.

\newpage
\section{UI}

\begin{figure*}[!h]
    \centering
    \includegraphics[width=0.9\linewidth]{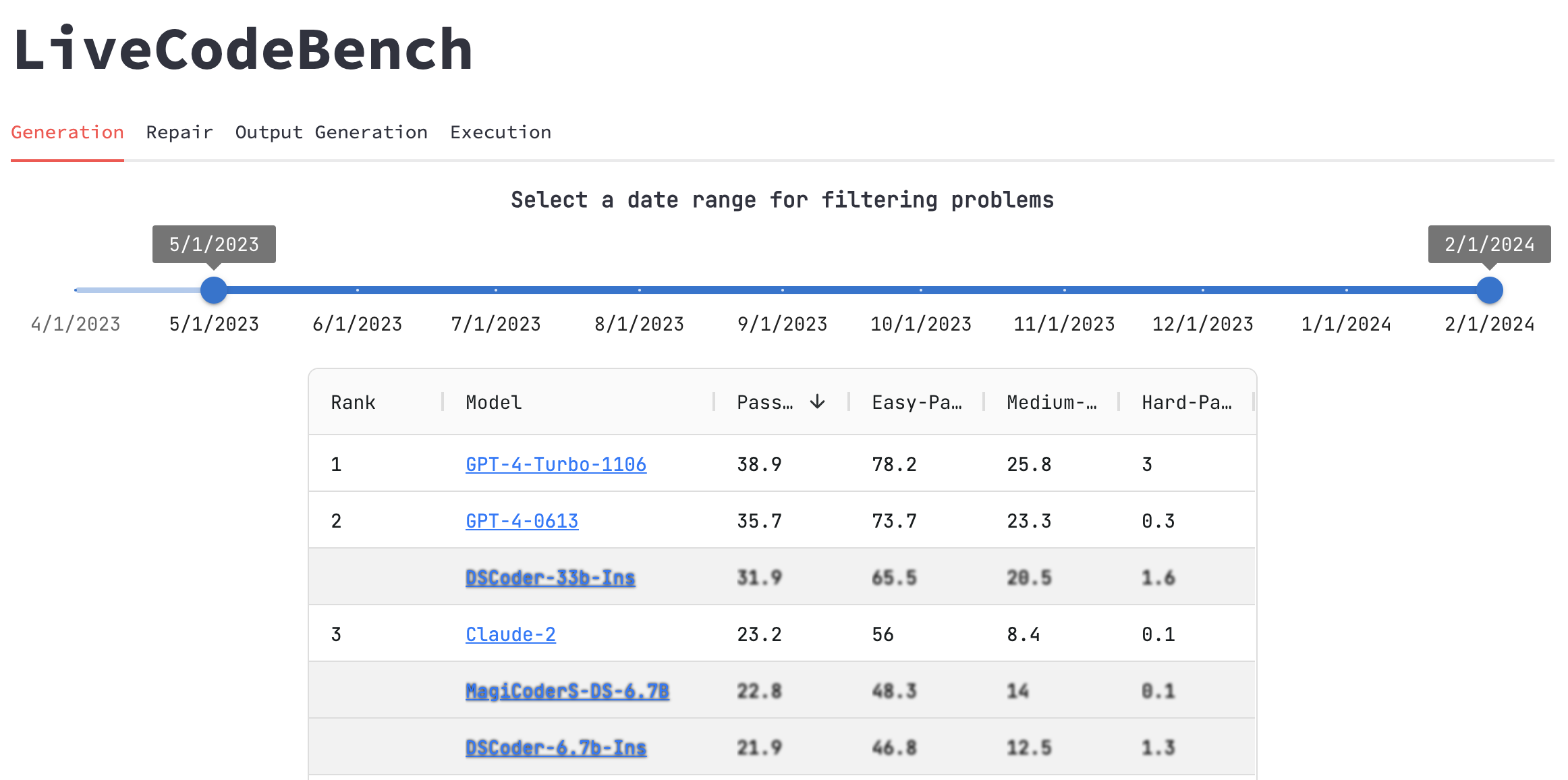}
    \includegraphics[width=0.9\linewidth]{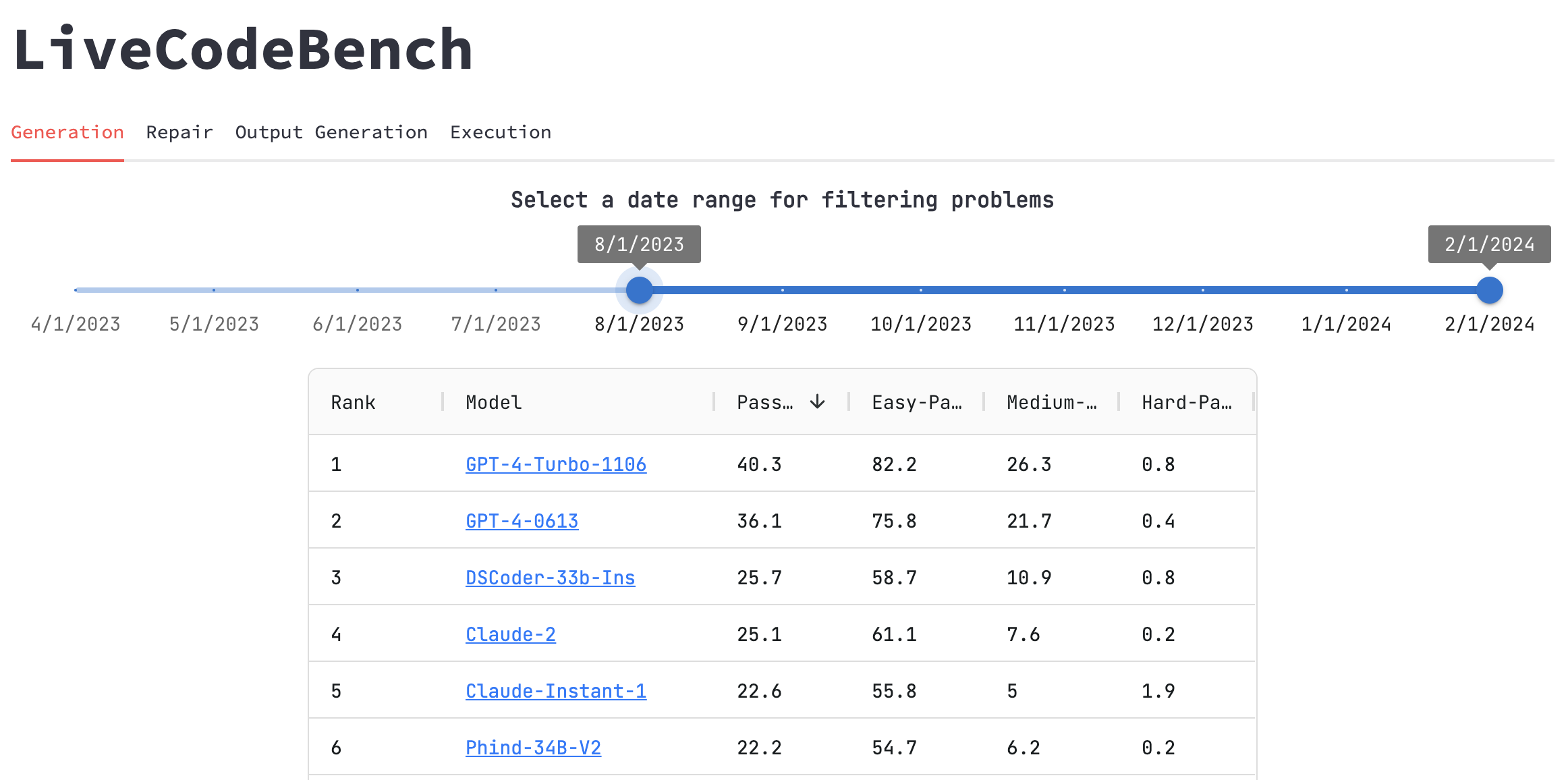}
    \caption{UI of \livecodebench{} showing two views -- May-Jan and Sep-Jan. The contaminated models are blurred and the performance difference is visible across the two views. The scroller on the top allows selecting different periods of time highlighting the live nature of the benchmark.}
    \label{app:fig:scrolling}
\end{figure*}
\newpage
\section{Experimental Setup}
\subsection{Models}
\label{app:subsec:models}
We describe the details of models considered in our study in Table~\ref{tab:models_list}.

\begin{longtable}{|p{0.3\linewidth}|p{0.22\linewidth}|p{0.15\linewidth}|p{0.28\linewidth}|}
    \textbf{Model ID} & \textbf{Short Name} & \textbf{Approximate Cutoff Date} & \textbf{Link} \\
deepseek-ai/deepseek-coder-33b-instruct & DSCoder-33b-Ins & 08/30/2023 & \href{https://huggingface.co/deepseek-ai/deepseek-coder-33b-instruct}{deepseek-coder-33b-instruct} \\ 
deepseek-ai/deepseek-coder-6.7b-instruct & DSCoder-6.7b-Ins & 08/30/2023 & \href{https://huggingface.co/deepseek-ai/deepseek-coder-6.7b-instruct}{deepseek-coder-6.7b-instruct} \\ 
deepseek-ai/deepseek-coder-1.3b-instruct & DSCoder-1.3b-Ins & 08/30/2023 & \href{https://huggingface.co/deepseek-ai/deepseek-coder-1.3b-instruct}{deepseek-coder-1.3b-instruct} \\ 
codellama/CodeLlama-70b-Instruct-hf & CodeLlama-70b-Ins & 01/01/2023 & \href{https://huggingface.co/codellama/CodeLlama-70b-Instruct-hf}{CodeLlama-70b-Instruct-hf} \\ 
openbmb/Eurus-70b-sft & Eurus-70B-SFT (n=1) & 01/01/2023 & \href{https://huggingface.co/openbmb/Eurus-70b-sft}{Eurus-70b-sft} \\ 
openbmb/Eurux-8x22b-nca & Eurux-8x22b-NCA (n=1) & 04/30/2023 & \href{https://huggingface.co/openbmb/Eurux-8x22b-nca}{Eurux-8x22b-nca} \\ 
codellama/CodeLlama-34b-Instruct-hf & CodeLlama-34b-Ins & 01/01/2023 & \href{https://huggingface.co/codellama/CodeLlama-34b-Instruct-hf}{CodeLlama-34b-Instruct-hf} \\ 
codellama/CodeLlama-13b-Instruct-hf & CodeLlama-13b-Ins & 01/01/2023 & \href{https://huggingface.co/codellama/CodeLlama-13b-Instruct-hf}{CodeLlama-13b-Instruct-hf} \\ 
codellama/CodeLlama-7b-Instruct-hf & CodeLlama-7b-Ins & 01/01/2023 & \href{https://huggingface.co/codellama/CodeLlama-7b-Instruct-hf}{CodeLlama-7b-Instruct-hf} \\ 
meta-llama/Meta-Llama-3-8B-Instruct & LLama3-8b-Ins & 01/01/2023 & \href{https://huggingface.co/meta-llama/Meta-Llama-3-8B-Instruct}{Meta-Llama-3-8B-Instruct} \\ 
meta-llama/Meta-Llama-3-70B-Instruct & LLama3-70b-Ins & 01/01/2023 & \href{https://huggingface.co/meta-llama/Meta-Llama-3-70B-Instruct}{Meta-Llama-3-70B-Instruct} \\ 
Phind/Phind-CodeLlama-34B-v2 & Phind-34B-V2 & 01/01/2023 & \href{https://huggingface.co/Phind/Phind-CodeLlama-34B-v2}{Phind-CodeLlama-34B-v2} \\ 
Smaug-2-72B & Smaug-2-72B & 01/01/2023 & \href{https://huggingface.co/abacusai/Smaug-2-72B}{Smaug-2-72B} \\ 
Qwen-1.5-72B-Chat & Qwen-1.5-72B-Chat & 01/01/2023 & \href{https://huggingface.co/qwen/Qwen1.5-72B-Chat}{Qwen-1.5-72B-Chat} \\ 
Qwen/CodeQwen1.5-7B & CodeQwen15-7B & 08/30/2023 & \href{https://huggingface.co/Qwen/CodeQwen1.5-7B}{CodeQwen1.5-7B} \\ 
Qwen/CodeQwen1.5-7B-Chat & CodeQwen15-7B-Chat & 08/30/2023 & \href{https://huggingface.co/Qwen/CodeQwen1.5-7B-Chat}{CodeQwen1.5-7B-Chat} \\ 
gpt-3.5-turbo-0301 & GPT-3.5-Turbo-0301 & 10/01/2021 & \href{https://openai.com/blog/new-models-and-developer-products-announced-at-devday}{gpt-3.5-turbo-0301} \\ 
gpt-3.5-turbo-0125 & GPT-3.5-Turbo-0125 & 10/01/2021 & \href{https://openai.com/blog/new-embedding-models-and-api-updates#:~:text=Other%20new%20models%20and%20lower%20pricing}{gpt-3.5-turbo-0125} \\ 
gpt-4-0613 & GPT-4-0613 & 10/01/2021 & \href{https://openai.com/blog/new-models-and-developer-products-announced-at-devday}{gpt-4-0613} \\ 
gpt-4-1106-preview & GPT-4-Turbo-1106 & 04/30/2023 & \href{https://openai.com/blog/new-models-and-developer-products-announced-at-devday}{gpt-4-1106-preview} \\ 
gpt-4-turbo-2024-04-09 & GPT-4-Turbo-2024-04-09 & 04/30/2023 & \href{https://openai.com/blog/new-models-and-developer-products-announced-at-devday}{gpt-4-turbo-2024-04-09} \\ 
gpt-4o-2024-05-13 & GPT-4O-2024-05-13 & 10/30/2023 & \href{https://openai.com/index/spring-update}{gpt-4o-2024-05-13} \\ 
claude-2 & Claude-2 & 12/31/2022 & \href{https://www.anthropic.com/index/claude-2}{claude-2} \\ 
claude-instant-1 & Claude-Instant-1 & 12/31/2022 & \href{https://www.anthropic.com/index/introducing-claude}{claude-instant-1} \\ 
claude-3-opus-20240229 & Claude-3-Opus & 04/30/2023 & \href{https://www.anthropic.com/claude}{claude-3-opus-20240229} \\ 
claude-3-sonnet-20240229 & Claude-3-Sonnet & 04/30/2023 & \href{https://www.anthropic.com/claude}{claude-3-sonnet-20240229} \\ 
claude-3-haiku-20240307 & Claude-3-Haiku & 04/30/2023 & \href{https://www.anthropic.com/claude}{claude-3-haiku-20240307} \\ 
codestral-latest & Codestral-Latest & 01/31/2024 & \href{https://mistral.ai/news/codestral/}{codestral-latest} \\ 
gemini-pro & Gemini-Pro & 04/30/2023 & \href{https://blog.google/technology/ai/gemini-api-developers-cloud}{gemini-pro} \\ 
gemini-1.5-pro-latest & Gemini-Pro-1.5-May & 04/30/2023 & \href{https://blog.google/technology/ai/gemini-api-developers-cloud}{gemini-1.5-pro-latest} \\ 
gemini-1.5-flash-latest & Gemini-Flash-1.5-May & 04/30/2023 & \href{https://blog.google/technology/ai/gemini-api-developers-cloud}{gemini-1.5-flash-latest} \\ 
ise-uiuc/Magicoder-S-DS-6.7B & MagiCoderS-DS-6.7B & 08/30/2023 & \href{https://huggingface.co/ise-uiuc/Magicoder-S-DS-6.7B}{Magicoder-S-DS-6.7B} \\ 
ise-uiuc/Magicoder-S-CL-7B & MagiCoderS-CL-7B & 01/01/2023 & \href{https://huggingface.co/ise-uiuc/Magicoder-S-CL-7B}{Magicoder-S-CL-7B} \\ 
bigcode/starcoder2-3b & StarCoder2-3b & 01/01/2023 & \href{https://huggingface.co/bigcode/starcoder2-3b}{starcoder2-3b} \\ 
bigcode/starcoder2-7b & StarCoder2-7b & 01/01/2023 & \href{https://huggingface.co/bigcode/starcoder2-7b}{starcoder2-7b} \\ 
bigcode/starcoder2-15b & StarCoder2-15b & 01/01/2023 & \href{https://huggingface.co/bigcode/starcoder2-15b}{starcoder2-15b} \\ 
codellama/CodeLlama-70b-hf & CodeLlama-70b-Base & 01/01/2023 & \href{https://huggingface.co/codellama/CodeLlama-70b-hf}{CodeLlama-70b-hf} \\ 
codellama/CodeLlama-34b-hf & CodeLlama-34b-Base & 01/01/2023 & \href{https://huggingface.co/codellama/CodeLlama-34b-hf}{CodeLlama-34b-hf} \\ 
codellama/CodeLlama-13b-hf & CodeLlama-13b-Base & 01/01/2023 & \href{https://huggingface.co/codellama/CodeLlama-13b-hf}{CodeLlama-13b-hf} \\ 
codellama/CodeLlama-7b-hf & CodeLlama-7b-Base & 01/01/2023 & \href{https://huggingface.co/codellama/CodeLlama-7b-hf}{CodeLlama-7b-hf} \\ 
deepseek-ai/deepseek-coder-33b-base & DSCoder-33b-Base & 08/30/2023 & \href{https://huggingface.co/deepseek-ai/deepseek-coder-33b-base}{deepseek-coder-33b-base} \\ 
deepseek-ai/deepseek-coder-6.7b-base & DSCoder-6.7b-Base & 08/30/2023 & \href{https://huggingface.co/deepseek-ai/deepseek-coder-6.7b-base}{deepseek-coder-6.7b-base} \\ 
deepseek-ai/deepseek-coder-1.3b-base & DSCoder-1.3b-Base & 08/30/2023 & \href{https://huggingface.co/deepseek-ai/deepseek-coder-1.3b-base}{deepseek-coder-1.3b-base} \\ 
google/codegemma-7b & CodeGemma-7b-Base & 01/01/2023 & \href{https://huggingface.co/google/codegemma-7b}{codegemma-7b} \\ 
google/codegemma-2b & CodeGemma-2b-Base & 01/01/2023 & \href{https://huggingface.co/google/codegemma-2b}{codegemma-2b} \\ 
google/gemma-7b & Gemma-7b-Base & 01/01/2023 & \href{https://huggingface.co/google/gemma-7b}{gemma-7b} \\ 
google/gemma-2b & Gemma-2b-Base & 01/01/2023 & \href{https://huggingface.co/google/gemma-2b}{gemma-2b} \\ 
meta-llama/Meta-Llama-3-70B & LLama3-70b-Base & 01/01/2023 & \href{https://huggingface.co/meta-llama/Meta-Llama-3-70B}{Meta-Llama-3-70B} \\ 
meta-llama/Meta-Llama-3-8B & LLama3-8b-Base & 01/01/2023 & \href{https://huggingface.co/meta-llama/Meta-Llama-3-8B}{Meta-Llama-3-8B} \\ 
mistral-large-latest & Mistral-Large & 01/01/2023 & \href{https://mistral.ai/news/mistral-large/}{mistral-large-latest} \\ 
open-mixtral-8x22b & Mixtral-8x22B-Ins & 01/01/2023 & \href{https://mistral.ai/news/mixtral-8x22b/}{open-mixtral-8x22b} \\ 
open-mixtral-8x7b & Mixtral-8x7B-Ins & 01/01/2023 & \href{https://mistral.ai/news/mixtral-8x7b/}{open-mixtral-8x7b} \\ 
m-a-p/OpenCodeInterpreter-DS-33B & OC-DS-33B & 08/30/2023 & \href{https://huggingface.co/m-a-p/OpenCodeInterpreter-DS-33B/}{OpenCodeInterpreter-DS-33B} \\ 
m-a-p/OpenCodeInterpreter-DS-6.7B & OC-DS-6.7B & 08/30/2023 & \href{https://huggingface.co/m-a-p/OpenCodeInterpreter-DS-6.7B/}{OpenCodeInterpreter-DS-6.7B} \\ 
m-a-p/OpenCodeInterpreter-DS-1.3B & OC-DS-1.3B & 08/30/2023 & \href{https://huggingface.co/m-a-p/OpenCodeInterpreter-DS-1.3B/}{OpenCodeInterpreter-DS-1.3B} \\ 
command-r & Command-R & 01/01/2023 & \href{https://docs.cohere.com/docs/models}{command-r} \\ 
command-r+ & Command-R+ & 01/01/2023 & \href{https://docs.cohere.com/docs/models}{command-r+} \\ 
    \caption{Language Models Overview}
    \label{tab:models_list}
    \end{longtable}

\subsection{Code Generation}
\label{appendix:codegen-prompts}
Below we provide the prompt format (with appropriate variants adding special tokens accommodating each instruct-tuned model) used for this scenario.

\begin{figure*}[h!]
\centering
\begin{tcolorbox}[enhanced,size=small,colback=black!5!white,colframe=RoyalBlue,flip title={interior hidden},title={Code Generation Prompt}]
\begin{lstlisting}[label={lst:execution-example-filtered-1}, captionpos=b, breaklines=true, language=Python]
You are an expert Python programmer. You will be given a question (problem specification) and will generate a correct Python program that matches the specification and passes all tests. You will NOT return anything except for the program

### Question:\n{question.question_content}


{ if question.starter_code }
 ### Format: {PromptConstants.FORMATTING_MESSAGE}

```python
{question.starter_code}
```
{ else }
### Format: {PromptConstants.FORMATTING_WITHOUT_STARTER_MESSAGE}

```python
# YOUR CODE HERE
```
{ endif }


### Answer: (use the provided format with backticks)
\end{lstlisting}
\end{tcolorbox}
\end{figure*}

\subsection{Self Repair}
\label{appendix:repair-prompts}
Below we provide the prompt format (with appropriate variants adding special tokens accommodating each instruct-tuned model) used for this scenario.

\begin{figure*}[h!]
\centering
\begin{tcolorbox}[enhanced,size=small,colback=black!5!white,colframe=RoyalBlue,flip title={interior hidden},title={Self Repair Error Feedback Pseudocode}]
\begin{lstlisting}[label={lst:execution-example-filtered-1}, captionpos=b, breaklines=true, language=Python]
{if check_result.result_status is "Wrong Answer"}
The above code is incorrect and does not pass the testcase.
Input: {wrong_testcase_input}
Output: {wrong_testcase_output}
Expected: {wrong_testcase_expected}


{elif check_result.result_status is "Time Limit Exceeded"}
The above code is incorrect and exceeds the time limit.
Input: {wrong_testcase_input}


{elif check_result.result_status is "Runtime Error"}
The above code is incorrect and has a runtime error.
Input: {wrong_testcase_input}
Error Message: {wrong_testcase_error_message}

{endif}
\end{lstlisting}
\end{tcolorbox}
\end{figure*}

\begin{figure*}[h!]
\centering
\begin{tcolorbox}[enhanced,size=small,colback=black!5!white,colframe=RoyalBlue,flip title={interior hidden},title={Self-Repair Prompt}]
\begin{lstlisting}[label={lst:execution-example-filtered-1}, captionpos=b, breaklines=true, language=Python]
You are a helpful programming assistant and an expert Python programmer. You are helping a user write a program to solve a problem. The user has written some code, but it has some errors and is not passing the tests. You will help the user by first giving a concise (at most 2-3 sentences) textual explanation of what is wrong with the code. After you have pointed out what is wrong with the code, you will then generate a fixed version of the program. You must put the entired fixed program within code delimiters only for once.

### Question:\n{question.question_content}

### Answer: ```python
{code.code_to_be_corrected}
```

### Format: {PromptConstants.FORMATTING_CHECK_ERROR_MESSAGE}

### Answer: (use the provided format with backticks)
\end{lstlisting}
\end{tcolorbox}
\end{figure*}

\subsection{Code Execution} 
\label{appendix:code-execution-prompts}
Below we provide the prompts for code execution with and without CoT. The prompts are modified versions of those from \citep{gu2024cruxeval} to fit the format of the samples in our benchmark.

\begin{figure*}[h!]
\centering
\begin{tcolorbox}[enhanced,size=small,colback=black!5!white,colframe=RoyalBlue,flip title={interior hidden},title={Code Execution Prompt}]
\begin{lstlisting}[label={lst:execution-example-filtered-1}, captionpos=b, breaklines=true, language=Python]
You are given a Python function and an assertion containing an input to the function. Complete the assertion with a literal (no unsimplified expressions, no function calls) containing the output when executing the provided code on the given input, even if the function is incorrect or incomplete. Do NOT output any extra information. Provide the full assertion with the correct output in [ANSWER] and [/ANSWER] tags, following the examples.

[PYTHON]
def repeatNumber(number : int) -> int:
    return number
assert repeatNumber(number = 17) == ??
[/PYTHON]
[ANSWER]
assert repeatNumber(number = 17) == 17
[/ANSWER]

[PYTHON]
def addCharacterA(string : str) -> str:
    return string + "a"
assert addCharacterA(string = "x9j") == ??
[/PYTHON]
[ANSWER]
assert addCharacterA(string = "x9j") == "x9ja"
[/ANSWER]

[PYTHON]
{code}
assert {input} == ??
[/PYTHON]
[ANSWER]
\end{lstlisting}
\end{tcolorbox}
\end{figure*}

\begin{figure*}[h!]
\centering
\begin{tcolorbox}[enhanced,size=small,colback=black!5!white,colframe=RoyalBlue,flip title={interior hidden},title={Code Execution Prompt with CoT}]
\begin{lstlisting}[label={lst:execution-example-filtered-1}, captionpos=b, breaklines=true, language=Python]
You are given a Python function and an assertion containing an input to the function. Complete the assertion with a literal (no unsimplified expressions, no function calls) containing the output when executing the provided code on the given input, even if the function is incorrect or incomplete. Do NOT output any extra information. Execute the program step by step before arriving at an answer, and provide the full assertion with the correct output in [ANSWER] and [/ANSWER] tags, following the examples.

[PYTHON]
def performOperation(s):
    s = s + s
    return "b" + s + "a"
assert performOperation(s = "hi") == ??
[/PYTHON]
[THOUGHT]
Let's execute the code step by step:

1. The function performOperation is defined, which takes a single argument s.
2. The function is called with the argument "hi", so within the function, s is initially "hi".
3. Inside the function, s is concatenated with itself, so s becomes "hihi".
4. The function then returns a new string that starts with "b", followed by the value of s (which is now "hihi"), and ends with "a".
5. The return value of the function is therefore "bhihia".
[/THOUGHT]
[ANSWER]
assert performOperation(s = "hi") == "bhihia"
[/ANSWER]

[PYTHON]
{code}
assert {input} == ??
[/PYTHON]
[THOUGHT]
\end{lstlisting}
\end{tcolorbox}
\end{figure*}

\subsection{Test Output Prediction}
\label{appendix:test-pred-prompts}
Below we provide the prompt format (with appropriate variants adding special tokens accommodating each instruct-tuned model) used for this scenario.


\begin{figure*}[h!]
\centering
\begin{tcolorbox}[enhanced,size=small,colback=black!5!white,colframe=RoyalBlue,flip title={interior hidden},title={Test Output Prediction Prompt}]
\begin{lstlisting}[label={lst:execution-example-filtered-1}, captionpos=b, breaklines=true, language=Python]
### Instruction: You are a helpful programming assistant and an expert Python programmer. You are helping a user to write a test case to help to check the correctness of the function. The user has written a input for the testcase. You will calculate the output of the testcase and write the whole assertion statement in the markdown code block with the correct output.

Problem:
{problem_statement}
 
Function:
```
{function_signature}
```
Please complete the following test case:

```
assert {function_name}({testcase_input}) == # TODO
```
### Response:

\end{lstlisting}
\end{tcolorbox}
\end{figure*}

\newpage
\section{Results}
\subsection{Contamination}
\begin{figure*}[!h]
    \includegraphics[width=0.49\linewidth]{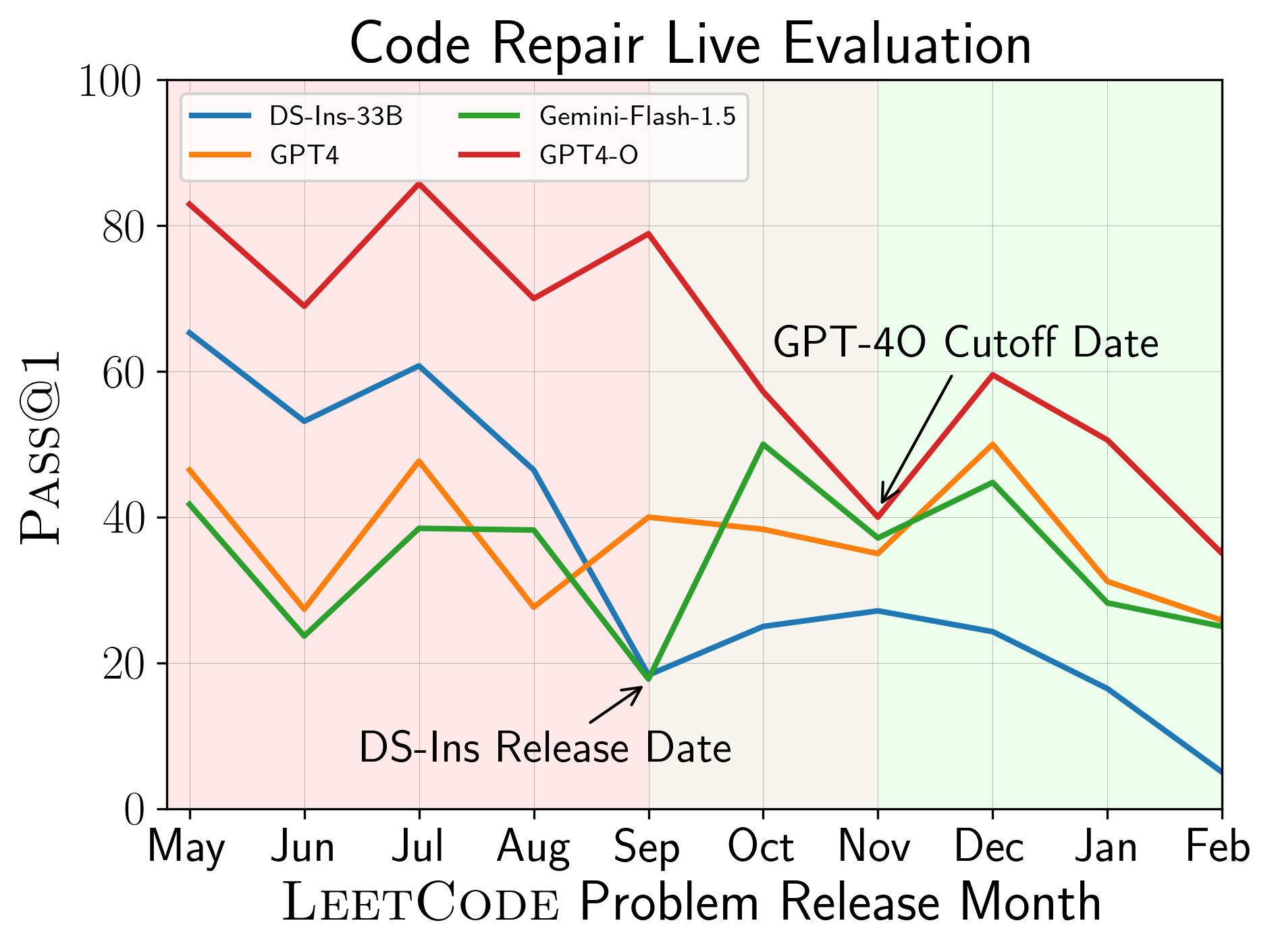}
    \hfil%
    \includegraphics[width=0.49\linewidth]{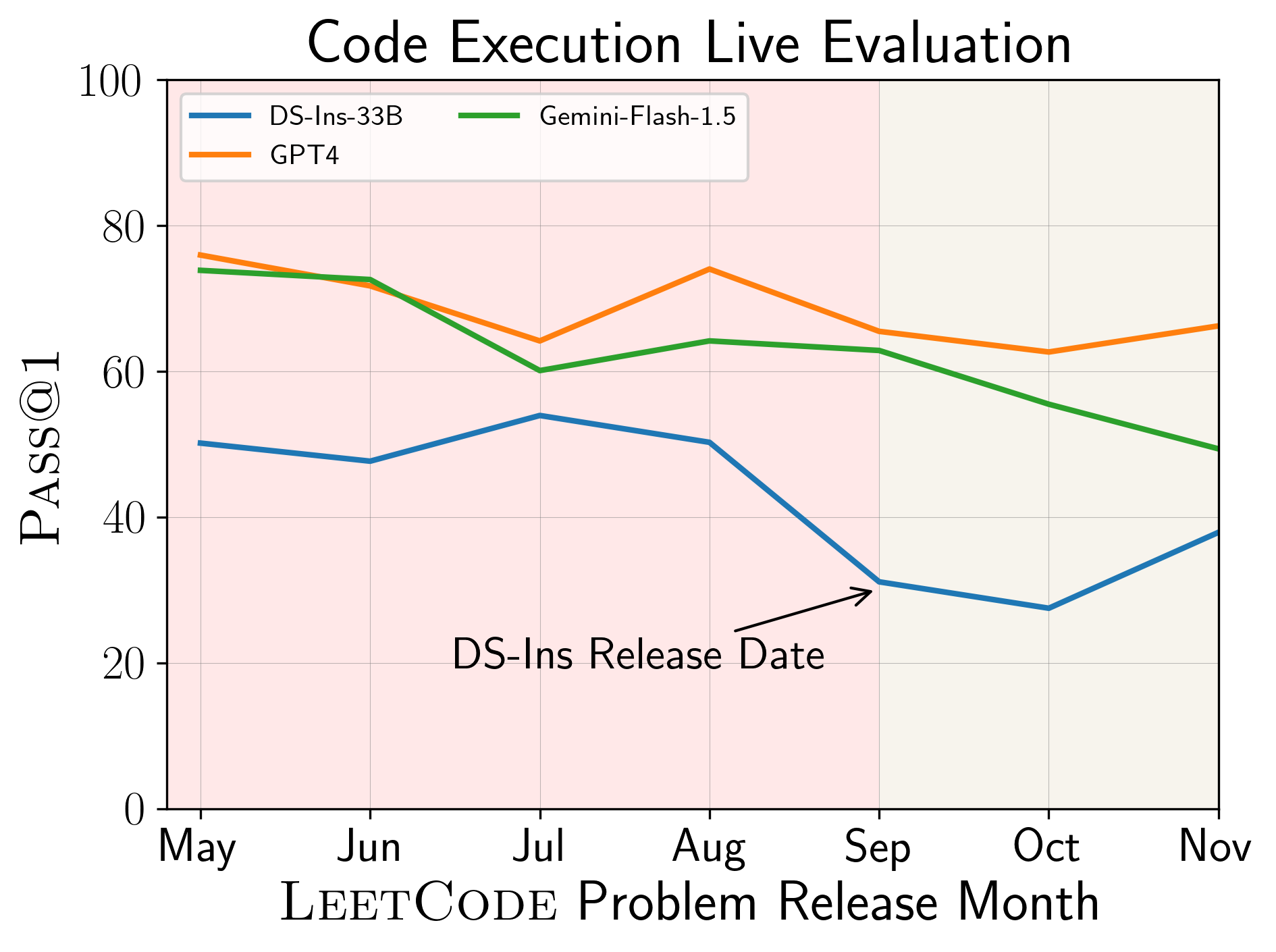}%

    \caption{Contamination in \deepseekcode{} models across self-repair and code execution (without \chainot{}) scenarios over time. Note that code execution currently runs between May and November}
    \label{fig:contamination_all_tasks_line}
\end{figure*}

Figure~\ref{fig:contamination_all_tasks_line} demonstrates contamination in \deepseek{} in self repair and test output prediction scenarios.

\begin{figure*}[!h]
\centering
    \includegraphics[width=0.49\linewidth]{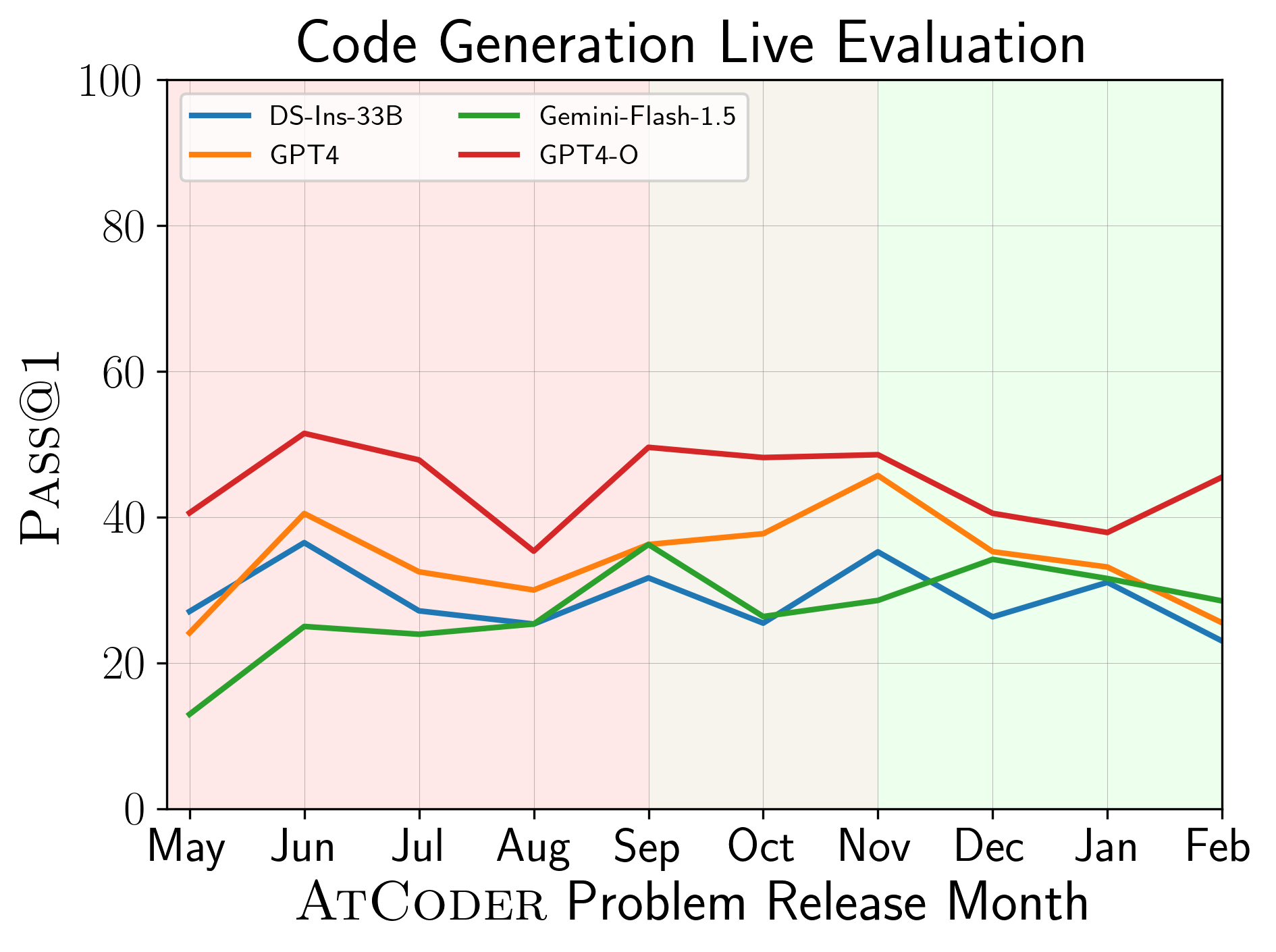}
        \caption{Performance on problems released over different months for \atcoder{}}
    \label{fig:atcoder_contamination}

\end{figure*}

\subsection{All Results}
Below we provide the tables comprising of results across different \livecodebench{} scenarios.
\begin{table*}[ht]
\centering
\begin{tabular}{lrrrr}
    \hline
Model Name & Easy & Medium & Hard & Total \\
    \hline
Claude-2 & 61.80& 4.90& 0.20& 22.30\\
Claude-3-Haiku & 63.00& 4.30& 1.10& 22.80\\
Claude-3-Opus & 78.80& 16.30& 3.20& 32.80\\
Claude-3-Sonnet & 67.60& 6.20& 1.10& 25.00\\
Claude-Instant-1 & 60.70& 4.30& 1.10& 22.10\\
CodeGemma-2b-Base & 18.30& 0.40& 0.00& 6.30\\
CodeGemma-7b-Base & 35.70& 2.60& 0.10& 12.80\\
CodeLlama-13b-Base & 24.60& 0.90& 0.00& 8.50\\
CodeLlama-13b-Ins & 36.60& 2.40& 0.00& 13.00\\
CodeLlama-34b-Base & 32.20& 1.80& 0.10& 11.40\\
CodeLlama-34b-Ins & 33.70& 2.40& 1.10& 12.40\\
CodeLlama-70b-Base & 15.80& 1.20& 0.00& 5.70\\
CodeLlama-70b-Ins & 7.80& 0.60& 0.00& 2.80\\
CodeLlama-7b-Base & 19.00& 0.40& 0.00& 6.50\\
CodeLlama-7b-Ins & 28.60& 2.50& 0.00& 10.40\\
CodeQwen15-7B & 40.40& 4.80& 0.00& 15.10\\
CodeQwen15-7B-Chat & 39.20& 13.10& 0.50& 17.60\\
Codestral-Latest & 69.00& 18.70& 0.90& 29.50\\
Command-R & 39.00& 3.60& 0.00& 14.20\\
Command-R+ & 56.60& 6.80& 0.00& 21.10\\
DSCoder-1.3b-Base & 17.30& 0.70& 0.00& 6.00\\
DSCoder-1.3b-Ins & 22.90& 1.50& 0.00& 8.10\\
DSCoder-33b-Base & 39.40& 2.30& 0.00& 13.90\\
DSCoder-33b-Ins & 55.60& 9.00& 0.70& 21.80\\
DSCoder-6.7b-Base & 34.30& 1.40& 0.10& 11.90\\
DSCoder-6.7b-Ins & 46.40& 5.80& 0.70& 17.60\\
GPT-3.5-Turbo-0125 & 56.80& 10.80& 0.10& 22.60\\
GPT-3.5-Turbo-0301 & 53.40& 8.80& 0.20& 20.80\\
GPT-4-0613 & 78.40& 21.20& 2.30& 33.90\\
GPT-4-Turbo-1106 & 84.40& 24.00& 0.50& 36.30\\
GPT-4-Turbo-2024-04-09 & 85.30& 33.00& 5.10& 41.10\\
GPT-4O-2024-05-13 & 88.30& 33.20& 4.20& 41.90\\
Gemini-Flash-1.5-May & 68.10& 12.60& 2.70& 27.80\\
Gemini-Pro-1.5-May & 76.00& 19.40& 3.50& 33.00\\
Gemma-7b-Base & 27.00& 0.90& 0.00& 9.30\\
LLama3-70b-Base & 52.20& 3.20& 0.60& 18.60\\
LLama3-70b-Ins & 60.70& 15.80& 1.40& 26.00\\
LLama3-8b-Base & 32.90& 1.50& 0.00& 11.50\\
LLama3-8b-Ins & 38.60& 3.50& 0.50& 14.20\\
MagiCoderS-DS-6.7B & 49.20& 7.50& 0.00& 18.90\\
Mistral-Large & 60.20& 10.90& 0.90& 24.00\\
Mixtral-8x22B-Ins & 59.80& 12.70& 0.00& 24.20\\
Mixtral-8x7B-Ins & 31.60& 2.60& 0.00& 11.40\\
OC-DS-1.3B & 11.30& 0.10& 0.00& 3.80\\
OC-DS-33B & 53.90& 5.10& 0.00& 19.70\\
OC-DS-6.7B & 46.30& 4.50& 0.00& 16.90\\
Phind-34B-V2 & 53.40& 4.70& 0.10& 19.40\\
StarCoder2-15b & 37.30& 2.20& 0.00& 13.20\\
StarCoder2-3b & 28.20& 0.70& 0.00& 9.60\\
StarCoder2-7b & 29.90& 1.20& 0.00& 10.40\\
    \hline
    \end{tabular}
\caption{Code Generation Performances}
\end{table*}

\begin{table*}[ht]
\centering
\begin{tabular}{lrrrr}
    \hline
Model Name & Easy & Medium & Hard & Total \\
    \hline
Claude-2 & 66.20& 10.30& 0.40& 25.60\\
Claude-3-Haiku & 66.50& 8.70& 2.50& 25.90\\
Claude-3-Opus & 83.10& 23.70& 6.70& 37.80\\
Claude-3-Sonnet & 72.60& 11.80& 2.20& 28.90\\
Claude-Instant-1 & 64.40& 7.10& 2.20& 24.60\\
CodeLlama-13b-Ins & 43.10& 3.00& 0.00& 15.30\\
CodeLlama-34b-Ins & 31.50& 3.50& 1.80& 12.30\\
CodeLlama-7b-Ins & 31.90& 3.10& 1.50& 12.10\\
Codestral-Latest & 72.50& 25.90& 3.30& 33.90\\
DSCoder-1.3b-Ins & 29.50& 2.10& 0.00& 10.60\\
DSCoder-33b-Ins & 60.70& 8.10& 1.50& 23.40\\
DSCoder-6.7b-Ins & 49.90& 5.70& 1.10& 18.90\\
GPT-3.5-Turbo-0125 & 59.30& 11.90& 0.50& 23.90\\
GPT-3.5-Turbo-0301 & 58.40& 11.60& 0.70& 23.60\\
GPT-4-0613 & 79.30& 25.00& 2.40& 35.60\\
GPT-4-Turbo-1106 & 86.90& 36.90& 4.00& 42.60\\
GPT-4-Turbo-2024-04-09 & 88.70& 39.70& 8.40& 45.60\\
GPT-4O-2024-05-13 & 92.60& 46.40& 8.20& 49.10\\
Gemini-Flash-1.5-May & 73.40& 16.40& 4.40& 31.40\\
Gemini-Pro & 53.80& 9.40& 0.20& 21.10\\
Gemini-Pro-1.5-April (n=1) & 71.80& 19.40& 5.50& 32.20\\
Gemini-Pro-1.5-May & 84.80& 30.10& 7.30& 40.70\\
LLama3-70b-Ins & 69.60& 19.00& 1.80& 30.10\\
LLama3-8b-Ins & 47.10& 6.10& 0.00& 17.70\\
MagiCoderS-CL-7B & 36.50& 3.10& 0.00& 13.20\\
MagiCoderS-DS-6.7B & 50.60& 8.60& 0.00& 19.70\\
Mistral-Large & 71.20& 15.60& 3.60& 30.10\\
OC-DS-1.3B & 20.00& 0.40& 0.00& 6.80\\
OC-DS-33B & 58.90& 7.20& 1.30& 22.50\\
OC-DS-6.7B & 50.90& 6.30& 0.20& 19.10\\
Phind-34B-V2 & 62.00& 6.50& 0.90& 23.10\\
    \hline
\end{tabular}
\caption{Self Repair Performances}
\end{table*}

\begin{table*}[ht]
\centering
\begin{tabular}{lr}
    \hline
Model Name & Pass@1 \\
    \hline
Claude-2 & 32.70\\
Claude-3-Haiku & 32.90\\
Claude-3-Opus & 58.70\\
Claude-3-Sonnet & 34.10\\
Claude-Instant-1 & 25.40\\
CodeLlama-13b-Ins & 24.40\\
CodeLlama-34b-Ins & 23.00\\
CodeLlama-70b-Ins & 16.10\\
CodeLlama-7b-Ins & 15.30\\
Codestral-Latest & 41.80\\
DSCoder-1.3b-Ins & 12.50\\
DSCoder-33b-Ins & 28.30\\
DSCoder-6.7b-Ins & 26.50\\
GPT-3.5-Turbo-0125 & 35.40\\
GPT-3.5-Turbo-0301 & 32.50\\
GPT-4-0613 & 52.90\\
GPT-4-Turbo-1106 & 55.70\\
GPT-4-Turbo-2024-04-09 & 66.10\\
GPT-4O-2024-05-13 & 68.90\\
Gemini-Flash-1.5-May & 38.10\\
Gemini-Pro & 29.50\\
Gemini-Pro-1.5-April (n=1) & 49.60\\
Gemini-Pro-1.5-May & 44.80\\
LLama3-70b-Ins & 41.40\\
LLama3-8b-Ins & 24.40\\
MagiCoderS-CL-7B & 21.30\\
MagiCoderS-DS-6.7B & 27.10\\
Mistral-Large & 46.50\\
Mixtral-8x22B-Ins & 44.70\\
Mixtral-8x7B-Ins & 31.80\\
OC-DS-1.3B & 7.80\\
OC-DS-33B & 11.30\\
OC-DS-6.7B & 18.30\\
Phind-34B-V2 & 27.20\\
    \hline
\end{tabular}
\caption{Test Output Prediction Performances}
\end{table*}

\begin{table*}[ht]
\centering
\begin{tabular}{lrr}
    \hline
Model Name & Pass@1 & Pass@1 (COT) \\
    \hline
Claude-2 & 31.50 & 43.80 \\
Claude-3-Haiku & 0.70 & 28.30 \\
Claude-3-Opus & 36.50 & 80.10 \\
Claude-3-Sonnet & 29.30 & 39.40 \\
Claude-Instant-1 & 20.00 & 34.80 \\
Cllama-13b-Ins & 23.50 & 14.10 \\
Cllama-34b-Ins & 28.90 & 24.50 \\
Cllama-7b-Ins & 20.60 & 14.20 \\
CodeLlama-70b-Ins & 31.20 & -1.00 \\
Codestral-Latest & 37.90 & 41.80 \\
DSCoder-1.3b-Base & 19.00 & 13.40 \\
DSCoder-1.3b-Ins & 18.10 & 17.00 \\
DSCoder-33b-Base & 29.90 & 29.10 \\
DSCoder-33b-Ins & 26.60 & 31.70 \\
DSCoder-6.7b-Base & 23.50 & 25.10 \\
DSCoder-6.7b-Ins & 23.10 & 23.80 \\
GPT-3.5-Turbo-0301 & 33.90 & 34.80 \\
GPT-4-0613 & 44.30 & 64.80 \\
GPT-4-Turbo-1106 & 40.50 & 83.60 \\
GPT-4-Turbo-2024-04-09 & 45.90 & 83.80 \\
GPT-4O-2024-05-13 & 39.10 & 91.00 \\
Gemini-Flash-1.5-May & 21.40 & 57.10 \\
Gemini-Pro & 27.70 & 37.40 \\
Gemini-Pro-1.5 (April) (n=1) & 30.30 & 64.40 \\
Gemini-Pro-1.5-May & 42.10 & 72.10 \\
LLama3-70b-Ins & 29.60 & 55.50 \\
LLama3-8b-Ins & 18.40 & 29.40 \\
MagiCoderS-CL-7B & 21.20 & -1.00 \\
MagiCoderS-DS-6.7B & 27.20 & -1.00 \\
Mistral-Large & 36.60 & 54.40 \\
Phind-34B-V2 & 26.90 & -1.00 \\
StarCoder & 20.30 & -1.00 \\
WCoder-34B-V1 & 28.40 & -1.00 \\
    \hline
\end{tabular}
\caption{Code Execution Performances}
\end{table*}

\begin{figure*}[!h]
    \centering
    \includegraphics[width=0.49\linewidth]{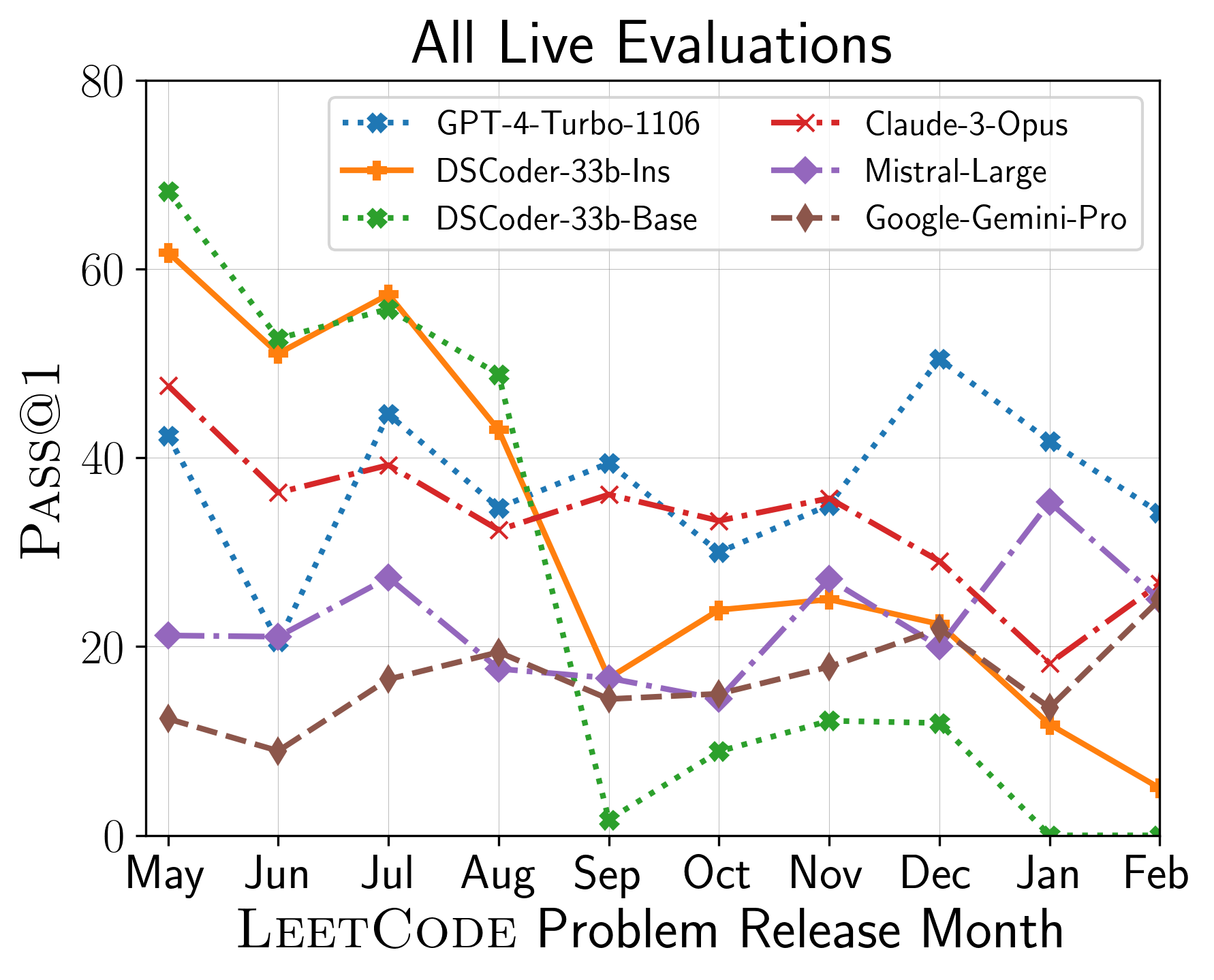}
    \includegraphics[width=0.49\linewidth]{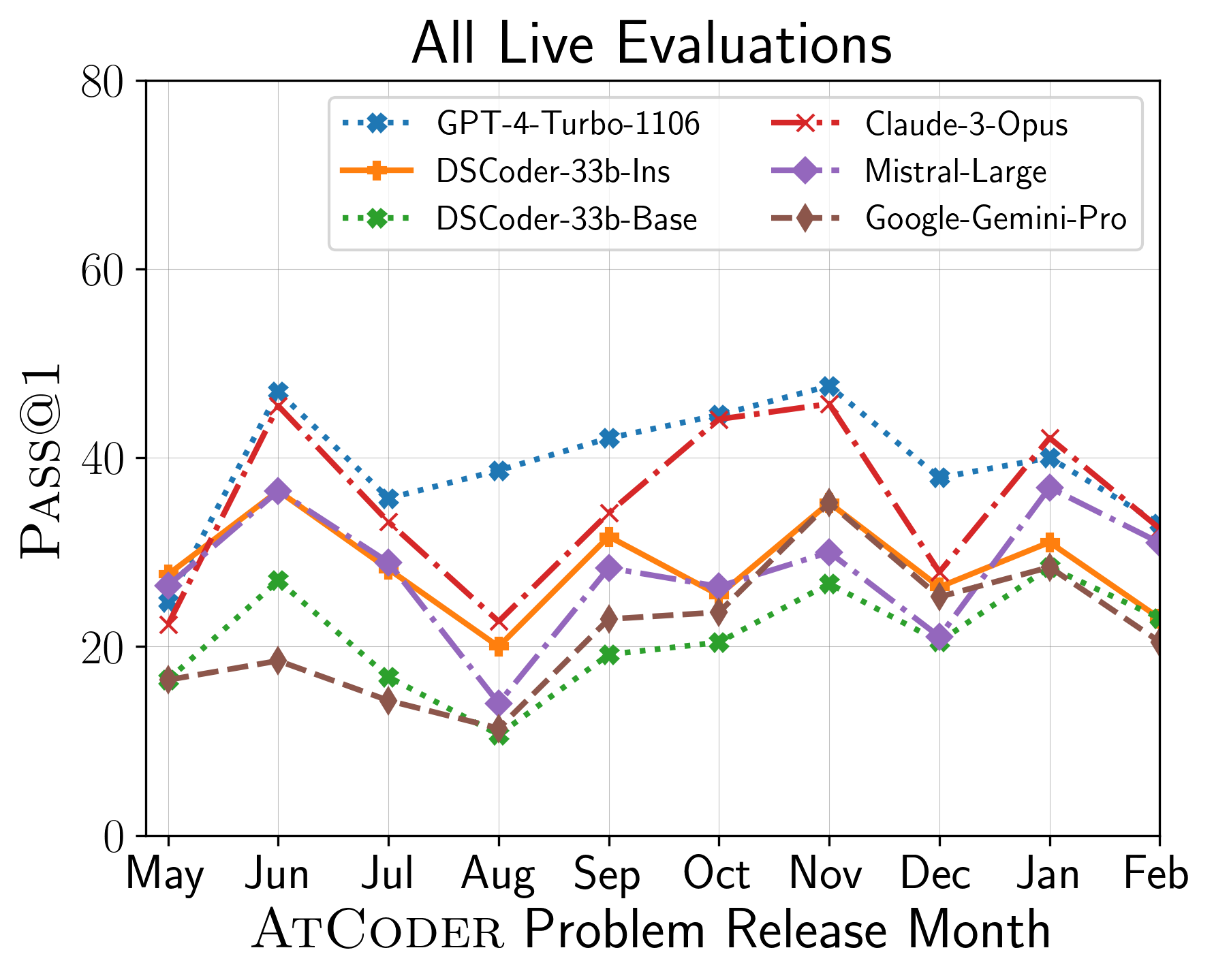}
    \caption{Live evaluation over time for various models on code generation scenario in \livecodebench{}. We consider many recently released models and do not find significant performance variations across months except for \deepseekcode{} models.}
    \label{app:fig:contamination_all_models}
\end{figure*}

\begin{figure*}[!h]
    \centering
    \includegraphics[width=0.49\linewidth]{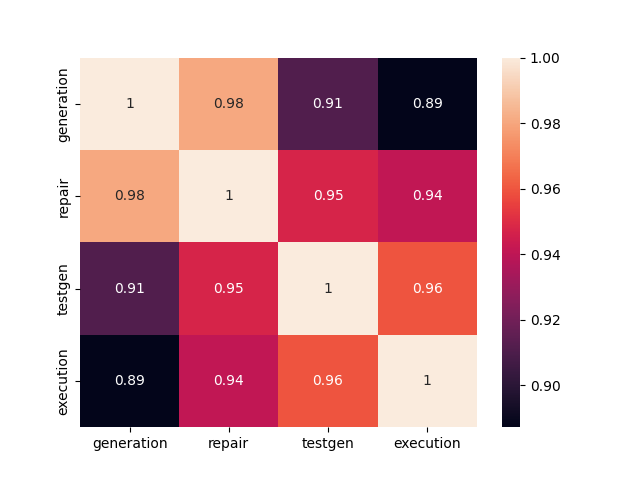}
    \caption{Correlations across different scenarios studied in \livecodebench{}}
    \label{app:fig:correlation_matrix}
\end{figure*}

\newpage
\section{Qualitative Examples}

\subsection{Code Execution}
We show 5 examples from the code execution task that GPT-4 (\texttt{gpt-4-1106-preview}) still struggles to execute, even with CoT.

\begin{figure*}[h!]
\centering
\begin{tcolorbox}[enhanced,size=small,colback=black!5!white,colframe=RoyalBlue,flip title={interior hidden},title={Mistake 1}]
\begin{lstlisting}[label={lst:execution-example-filtered-1}, captionpos=b, breaklines=true, language=Python]
def countWays(nums: List[int]) -> int:
    nums.sort()
    n = len(nums)
    ans = 0
    for i in range(n + 1):
        if i and nums[i-1] >= i: continue
        if i < n and nums[i] <= i: continue
        ans += 1
    return ans
assert countWays(nums = [6, 0, 3, 3, 6, 7, 2, 7]) == 3
# GPT-4 + CoT Outputs: 1, 2, 4, 5
\end{lstlisting}
\end{tcolorbox}
\end{figure*}

\begin{figure*}[h!]
\centering
\begin{tcolorbox}[enhanced,size=small,colback=black!5!white,colframe=RoyalBlue,flip title={interior hidden},title={Mistake 2}]
\begin{lstlisting}[label={lst:execution-example-filtered-1}, captionpos=b, breaklines=true, language=Python]
def minimumCoins(prices: List[int]) -> int:
    
    @cache
    def dfs(i, free_until):
        if i >= len(prices):
            return 0
        
        res = prices[i] + dfs(i + 1, min(len(prices) - 1, i + i + 1))
        
        if free_until >= i:
            res = min(res, dfs(i + 1, free_until))
            
        return res
        
    dfs.cache_clear()
    return dfs(0, -1)
assert minimumCoins(prices = [3, 1, 2]) == 4
# GPT-4 + CoT Outputs: 1, 3, 5, 6
\end{lstlisting}
\end{tcolorbox}
\end{figure*}

\begin{figure*}[h!]
\centering
\begin{tcolorbox}[enhanced,size=small,colback=black!5!white,colframe=RoyalBlue,flip title={interior hidden},title={Mistake 3}]
\begin{lstlisting}[label={lst:execution-example-filtered-1}, captionpos=b, breaklines=true, language=Python]
def sortVowels(s: str) -> str:
    q = deque(sorted((ch for ch in s if vowel(ch))))
    res = []
    for ch in s:
        if vowel(ch):
            res.append(q.popleft())
        else:
            res.append(ch)
    return ''.join(res)
assert sortVowels(s = 'lEetcOde') == 'lEOtcede'
# GPT-4 + CoT Outputs: "leetecode", "lEetecOde", "leetcede", "leetcEde", "leetcOde"
\end{lstlisting}
\end{tcolorbox}
\end{figure*}

\begin{figure*}[h!]
\centering
\begin{tcolorbox}[enhanced,size=small,colback=black!5!white,colframe=RoyalBlue,flip title={interior hidden},title={Mistake 4}]
\begin{lstlisting}[label={lst:execution-example-filtered-1}, captionpos=b, breaklines=true, language=Python]
def relocateMarbles(nums: List[int], moveFrom: List[int], moveTo: List[int]) -> List[int]:
    
    nums = sorted(list(set(nums)))
    dd = {}
    for item in nums:
        dd[item] = 1
    for a,b in zip(moveFrom, moveTo):
        del dd[a]
        dd[b] = 1
    ll = dd.keys()
    return sorted(ll)
assert relocateMarbles(nums = [1, 6, 7, 8], moveFrom = [1, 7, 2], moveTo = [2, 9, 5]) == [5, 6, 8, 9]
# GPT-4 + CoT Outputs: [2, 6, 8, 9], [2, 5, 6, 8, 9], KeyError
\end{lstlisting}
\end{tcolorbox}
\end{figure*}

\begin{figure*}[h!]
\centering
\begin{tcolorbox}[enhanced,size=small,colback=black!5!white,colframe=RoyalBlue,flip title={interior hidden},title={Mistake 5}]
\begin{lstlisting}[label={lst:execution-example-filtered-1}, captionpos=b, breaklines=true, language=Python]
def minimumSum(nums: List[int]) -> int:
    left, right, ans = [inf], [inf], inf
    for num in nums:
        left.append(min(left[-1], num))
    for num in nums[::-1]:
        right.append(min(right[-1], num))
    right.reverse()
    for i, num in enumerate(nums):
        if left[i] < num and right[i + 1] < num:
            ans = min(ans, num + left[i] + right[i + 1])
    return ans if ans < inf else -1
assert minimumSum(nums = [6, 5, 4, 3, 4, 5]) == -1
# GPT-4 + CoT Outputs: 10, 11, 12
\end{lstlisting}
\end{tcolorbox}
\end{figure*}

\end{document}